\documentclass[%
 aip,
 amsmath,amssymb,
 reprint,%
]{revtex4-1}

\usepackage[colorlinks=true, linkcolor=blue, citecolor=blue, urlcolor=blue]{hyperref}
\usepackage{graphicx}% Include figure files
\usepackage{xcolor}
\usepackage{dcolumn}% Align table columns on decimal point
\usepackage{bm}% bold math
\usepackage[utf8]{inputenc}
\usepackage[T1]{fontenc}
\usepackage{mathptmx}
\usepackage{etoolbox}
\usepackage{booktabs}
\usepackage{float}
\usepackage{amssymb}

\makeatletter
\def\@email#1#2{%
	\endgroup
	\patchcmd{\titleblock@produce}
	{\frontmatter@RRAPformat}
	{\frontmatter@RRAPformat{\produce@RRAP{*#1\href{mailto:#2}{#2}}}\frontmatter@RRAPformat}
	{}{}
}%

\renewcommand\frontmatter@title@format{%
	\fontsize{15.5pt}{18.5pt}\selectfont
	\bfseries
	\parindent\z@
	\parskip\z@skip
	\centering
}

\renewcommand\frontmatter@authorformat{%
	\large
	\centering
	\parindent\z@\relax
}

\renewcommand\frontmatter@affiliationfont{%
	\small
	\centering
	\parindent\z@\relax
}

\renewcommand\frontmatter@RRAP@format{%
	\small
	\centering
	\leftskip\z@\relax
	\parindent\z@\relax
}

\renewcommand\frontmatter@abstractfont{%
	\small
	\parindent1em\relax
}

\makeatother

\begin{document}

\preprint{AIP/123-QED}

\title{Tunable high-charge relativistic electron beams via direct laser acceleration in hohlraum-preheated foam targets\vspace{3pt}}

\author{Ziyao Wang}
\affiliation{MOE Key Laboratory for Nonequilibrium Synthesis and Modulation of Condensed Matter, \hbox{School of Physics, Xi'an Jiaotong University, Xi'an 710049, China}}
\affiliation{\hbox{National Key Laboratory of Plasma Physics, Laser Fusion Research Center, } \hbox{China Academy of Engineering Physics, Mianyang 621900, China}} 

\author{\hbox{Jieru Ren}}
\altaffiliation{Author to whom correspondence should be addressed:\\ \href{mailto:renjieru@xjtu.edu.cn}{renjieru@xjtu.edu.cn}, \href{mailto:dzgzju@163.com}{dzgzju@163.com}, \href{mailto:wenqingwzx@xjtu.edu.cn}{wenqingwzx@xjtu.edu.cn} 
}
\affiliation{MOE Key Laboratory for Nonequilibrium Synthesis and Modulation of Condensed Matter, \hbox{School of Physics, Xi'an Jiaotong University, Xi'an 710049, China}}

\author{\hbox{Zhigang Deng}}
\altaffiliation{Author to whom correspondence should be addressed:\\ \href{mailto:renjieru@xjtu.edu.cn}{renjieru@xjtu.edu.cn}, \href{mailto:dzgzju@163.com}{dzgzju@163.com}, \href{mailto:wenqingwzx@xjtu.edu.cn}{wenqingwzx@xjtu.edu.cn} 
}
\affiliation{\hbox{National Key Laboratory of Plasma Physics, Laser Fusion Research Center, } \hbox{China Academy of Engineering Physics, Mianyang 621900, China}} 

\author{\hbox{Wenqing Wei}}
\altaffiliation{Author to whom correspondence should be addressed:\\ \href{mailto:renjieru@xjtu.edu.cn}{renjieru@xjtu.edu.cn}, \href{mailto:dzgzju@163.com}{dzgzju@163.com}, \href{mailto:wenqingwzx@xjtu.edu.cn}{wenqingwzx@xjtu.edu.cn} 
}
\affiliation{MOE Key Laboratory for Nonequilibrium Synthesis and Modulation of Condensed Matter, \hbox{School of Physics, Xi'an Jiaotong University, Xi'an 710049, China}}

\author{\hbox{Wei Qi}}
\affiliation{\hbox{National Key Laboratory of Plasma Physics, Laser Fusion Research Center, } \hbox{China Academy of Engineering Physics, Mianyang 621900, China}} 

\author{\hbox{Olga N. Rosmej}}
\affiliation{\hbox{GSI Helmholtzzentrum für Schwerionenforschung, Darmstadt, 64291, Germany}} 

\author{\hbox{Nikolay E. Andreev}}
\affiliation{\hbox{Joint Institute for High Temperatures, Russian Academy of Sciences, Moscow, 125412, Russia}}

\author{\hbox{Sergey Yu. Gus'kov}}
\affiliation{\hbox{P.N. Lebedev Physical Institute, Russian Academy of Sciences, Moscow, 119991, Russia}} 

\author{\hbox{Rafael Yakhin}}
\affiliation{\hbox{P.N. Lebedev Physical Institute, Russian Academy of Sciences, Moscow, 119991, Russia}} 

\author{\hbox{Yifang Gao}}
\affiliation{MOE Key Laboratory for Nonequilibrium Synthesis and Modulation of Condensed Matter, \hbox{School of Physics, Xi'an Jiaotong University, Xi'an 710049, China}}

\author{\hbox{Bubo Ma}}
\affiliation{MOE Key Laboratory for Nonequilibrium Synthesis and Modulation of Condensed Matter, \hbox{School of Physics, Xi'an Jiaotong University, Xi'an 710049, China}}

\author{\hbox{Mingzhe Yang}}
\affiliation{MOE Key Laboratory for Nonequilibrium Synthesis and Modulation of Condensed Matter, \hbox{School of Physics, Xi'an Jiaotong University, Xi'an 710049, China}}

\author{\hbox{Shizheng Zhang}}
\affiliation{MOE Key Laboratory for Nonequilibrium Synthesis and Modulation of Condensed Matter, \hbox{School of Physics, Xi'an Jiaotong University, Xi'an 710049, China}}

\author{\hbox{Xuyang Luo}}
\affiliation{MOE Key Laboratory for Nonequilibrium Synthesis and Modulation of Condensed Matter, \hbox{School of Physics, Xi'an Jiaotong University, Xi'an 710049, China}}

\author{\hbox{Dieter H.H. Hoffmann}}
\affiliation{MOE Key Laboratory for Nonequilibrium Synthesis and Modulation of Condensed Matter, \hbox{School of Physics, Xi'an Jiaotong University, Xi'an 710049, China}} 

\author{\hbox{Peng Zhou}}
\affiliation{MOE Key Laboratory for Nonequilibrium Synthesis and Modulation of Condensed Matter, \hbox{School of Physics, Xi'an Jiaotong University, Xi'an 710049, China}} 

\author{\hbox{Ke Jiang}}
\affiliation{\hbox{Shenzhen Key Laboratory of Ultraintense Laser and Advanced Material Technology, Center for Intense Laser Application Technology,} \hbox{and College of Engineering Physics, Shenzhen Technology University, Shenzhen 518118, China}} 

\author{\hbox{Taiwu Huang}}
\affiliation{\hbox{Shenzhen Key Laboratory of Ultraintense Laser and Advanced Material Technology, Center for Intense Laser Application Technology,} \hbox{and College of Engineering Physics, Shenzhen Technology University, Shenzhen 518118, China}} 

\author{\hbox{Bo Cui}}
\affiliation{\hbox{National Key Laboratory of Plasma Physics, Laser Fusion Research Center, } \hbox{China Academy of Engineering Physics, Mianyang 621900, China}} 

\author{\hbox{Weiwu Wang}}
\affiliation{\hbox{National Key Laboratory of Plasma Physics, Laser Fusion Research Center, } \hbox{China Academy of Engineering Physics, Mianyang 621900, China}} 

\author{\hbox{Shaoyi Wang}}
\affiliation{\hbox{National Key Laboratory of Plasma Physics, Laser Fusion Research Center, } \hbox{China Academy of Engineering Physics, Mianyang 621900, China}} 

\author{\hbox{Quanping Fan}}
\affiliation{\hbox{National Key Laboratory of Plasma Physics, Laser Fusion Research Center, } \hbox{China Academy of Engineering Physics, Mianyang 621900, China}} 

\author{\hbox{Zhurong Cao}}
\affiliation{\hbox{National Key Laboratory of Plasma Physics, Laser Fusion Research Center, } \hbox{China Academy of Engineering Physics, Mianyang 621900, China}} 

\author{\hbox{Sixin Wu}}
\affiliation{\hbox{National Key Laboratory of Plasma Physics, Laser Fusion Research Center, } \hbox{China Academy of Engineering Physics, Mianyang 621900, China}} 

\author{\hbox{Yue Yang}}
\affiliation{\hbox{National Key Laboratory of Plasma Physics, Laser Fusion Research Center, } \hbox{China Academy of Engineering Physics, Mianyang 621900, China}} 

\author{\hbox{Leifeng Cao}}
\affiliation{\hbox{Shenzhen Key Laboratory of Ultraintense Laser and Advanced Material Technology, Center for Intense Laser Application Technology,} \hbox{and College of Engineering Physics, Shenzhen Technology University, Shenzhen 518118, China}} 

\author{\hbox{Yuqiu Gu}}
\affiliation{\hbox{National Key Laboratory of Plasma Physics, Laser Fusion Research Center, } \hbox{China Academy of Engineering Physics, Mianyang 621900, China}} 

\author{\hbox{Yuchi Wu}}
\affiliation{\hbox{National Key Laboratory of Plasma Physics, Laser Fusion Research Center, } \hbox{China Academy of Engineering Physics, Mianyang 621900, China}} 

\author{\hbox{Weimin Zhou}}
\affiliation{\hbox{National Key Laboratory of Plasma Physics, Laser Fusion Research Center, } \hbox{China Academy of Engineering Physics, Mianyang 621900, China}}

\author{\hbox{Zongqing Zhao}}
\affiliation{\hbox{National Key Laboratory of Plasma Physics, Laser Fusion Research Center, } \hbox{China Academy of Engineering Physics, Mianyang 621900, China}} 

\author{\hbox{Yongtao Zhao}\vspace{7pt}}
\affiliation{MOE Key Laboratory for Nonequilibrium Synthesis and Modulation of Condensed Matter, \hbox{School of Physics, Xi'an Jiaotong University, Xi'an 710049, China}}

\begin{abstract}

Direct laser acceleration (DLA) in near-critical-density (NCD) plasmas can  efficiently generate high-charge relativistic electron beams, yet beam parameters depend critically on precise plasma state manipulation. Solid-ablation NCD plasmas evolve rapidly, posing severe controllability challenges. We produce NCD plasma via indirectly heating foam targets with ns laser driven hohlraum soft X-ray. Electrons are generated through irradiating the plasma with another picosecond laser. Tuning the laser pulse delay $\tau$ enables control of plasma profiles and beam parameters.
Experiments show that when the foam is heated ($\tau$ = 6 ns, 9 ns), the beam exhibits $T \sim 13$ MeV effective temperature, $E_k \sim 80$ MeV cutoff energy, and hundreds of nC/sr charge for $E_k > 7.5$ MeV. These  values are significantly higher than those from solid-foil ($T$ $\sim$ 2.7 MeV, $E_k$ $\sim$ 20 MeV, $Q$ $\sim$ 9 nC/sr) and cold-foam ($T$ $\sim$ 12 MeV, $E_k$ $\sim$ 50 MeV, $Q$ $\sim$ 5 nC/sr) interactions. At a longer delay of $\tau$ = 15 ns, the charge increases further while the temperature decreases, and at a shorter delay of $\tau$ = 3 ns, both temperature and charge are lower.
3D PIC simulations link these observations to the interplay between the microstructure of the cold foam and the evolving plasma density profile at different delay times, which together determine the beam charge, effective temperature, and divergence. The finding provides a routine to generate and tailor the relativistic electron beams, which is essential for designing laser-driven electron sources for high energy density physics and photonuclear reaction applications.

\end{abstract}

\maketitle

\section{INTRODUCTION}

Laser-driven relativistic electron beams and their associated secondary radiation sources (such as X/$\gamma$ rays, protons, neutrons) feature ultrashort pulse duration and ultrahigh current density\cite{pomerantzUltrashortPulsedNeutron2014,guntherForwardlookingInsightsLasergenerated2022,fengHighefficiencyNeutronSource2020,tavanaUltrahighEfficiencyBremsstrahlung2023,cikhardtCharacterizationBrightBetatron2024,gyrdymovHighbrightnessBetatronEmission2024c,tangtartharakulCollimatedGrayEmission2025,tavanaUltrahighFluxDirect2026}. These characteristics make them highly valuable for a wide range of applications, including laboratory astrophysics\cite{zhangStudiesHighEnergy2016,takabeRecentProgressLaboratory2021}, medical isotope production\cite{maPhotonuclearProductionMedical2019,caoExperimentalStudyMedical2023a,pangPangProductionMedicalRadioisotopes512026a,yangExpressDiagnosticIntense2026a}, FLASH radiobiology\cite{flaccoLaserDrivenFLASH2025,gyrdymovUltraintensePulsedSource2026}, and ultrafast imaging\cite{kneipBrightSpatiallyCoherent2010,courtoisHighresolutionMultiMeVXray2011,paganoSourceSizeRays2024}. The efficient generation and stable control of high-charge electron beams have long been a critical goal of the high-energy-density (HED) physics community.

Laser wakefield acceleration (LWFA)\citealp{tajimaLaserElectronAccelerator1979a} in low-density plasma driven by femtosecond lasers can produce quasimonoenergetic electron beams with energies up to 7.8 GeV\cite{gonsalvesPetawattLaserGuiding2019}. However, due to the limited charge capacity of the plasma wake, the beam charge is typically restricted to 10--100 pC\cite{wangQuasimonoenergeticLaserplasmaAcceleration2013a,leemansMultiGeVElectronBeams2014a,kimStableMultiGeVElectron2017a}. In contrast, direct laser acceleration (DLA)\cite{pukhovParticleAccelerationRelativistic1999,gahnMultiMeVElectronBeam1999a} driven by picosecond lasers in near-critical-density (NCD) plasma---which relies on the combined effect of the laser electric field and the quasi-static electromagnetic fields formed in plasma channels---can generate a relativistic electron beam with a charge in the range of hundreds of nanocoulombs (nC)\cite{willingaleSurfaceWavesElectron2013a,pugachevAccelerationElectronsAction2016a,maUltrahighchargeElectronBeams2018,rosmejInteractionRelativisticallyIntense2019a,rosmejHighcurrentLaserdrivenBeams2020a,rosmejBrightBetatronRadiation2021a,husseinOptimisationDirectLaser2021a}. This makes DLA a promising approach for producing high-charge relativistic electron beams. 

Numerical studies have shown that the DLA process is highly sensitive to initial laser--target parameters, such as laser focusing conditions\cite{tangInfluenceLaserFocusing2024a}, the shape of the preformed plasma channel\cite{huangRelativisticLaserHosing2017a}, the atomic number of the target material \cite{cohenUndepletedDirectLaser2024a}, and the plasma density distribution\cite{babjakDirectLaserAcceleration2024b}. It was experimentally demonstrated that NCD plasmas can significantly enhance beam charge and cut-off energy\cite{husseinOptimisationDirectLaser2021a,kneipObservationSynchrotronRadiation2008a}. However, in those experiments, the NCD plasmas were produced either by solid-target ablation\cite{husseinOptimisationDirectLaser2021a} or by highly supersonic gas nozzles\cite{kneipObservationSynchrotronRadiation2008a}. Such plasmas usually suffer from rapid evolution and steep density gradients, which severely limits controllability and reproducibility\cite{rosmejHydrodynamicRadiativeProperties2015,chenDensityTemperatureCharacterization2016}. Low-density porous foam targets offer an alternative route to generate NCD plasmas\cite{rosmejHydrodynamicRadiativeProperties2015}. They were initially used in inertial confinement fusion (ICF) experiments owing to their ability to enable uniform laser energy deposition\cite{dunneEvaluationFoamBuffer1995}.  Rosmej et al. utilized nanosecond pre-pulses to directly heat a foam target to an NCD plasma state and demonstrated improved laser-to-electron energy conversion efficiency\cite{rosmejInteractionRelativisticallyIntense2019a,rosmejHighcurrentLaserdrivenBeams2020a,rosmejBrightBetatronRadiation2021a,rosmejAdvancedPlasmaTarget2025}.

In this work, we propose to generate NCD plasma by indirectly heating a foam target with soft X-rays emitted from a nanosecond-laser-driven hohlraum. This plasma is then used to produce high-charge relativistic electron beams via irradiation with a subsequent picosecond laser pulse. Previous studies have demonstrated that, under such indirect heating, the foam plasma expands slowly over nanosecond timescales and can survive for tens of nanoseconds, resulting in time-dependent density profiles with expansion rates much lower than those of solid targets\cite{rosmejHeatingLowdensityCHOfoam2011,rosmejHydrodynamicRadiativeProperties2015}. This approach allows fine control of the plasma density by choosing an appropriate foam density, and enables manipulation of the density profile by varying the time delay between the nanosecond and picosecond lasers. Our experimental results show that compared with either a solid foil or a cold foam target, the electron beam generated in the NCD plasma exhibits significantly higher charge and temperature. Moreover, the beam parameters vary systematically with the time delay. These observations are interpreted with the aid of three-dimensional particle-in-cell (3D-PIC) simulations.

\begin{figure}[b]
	\centering
	\includegraphics[width=\linewidth]{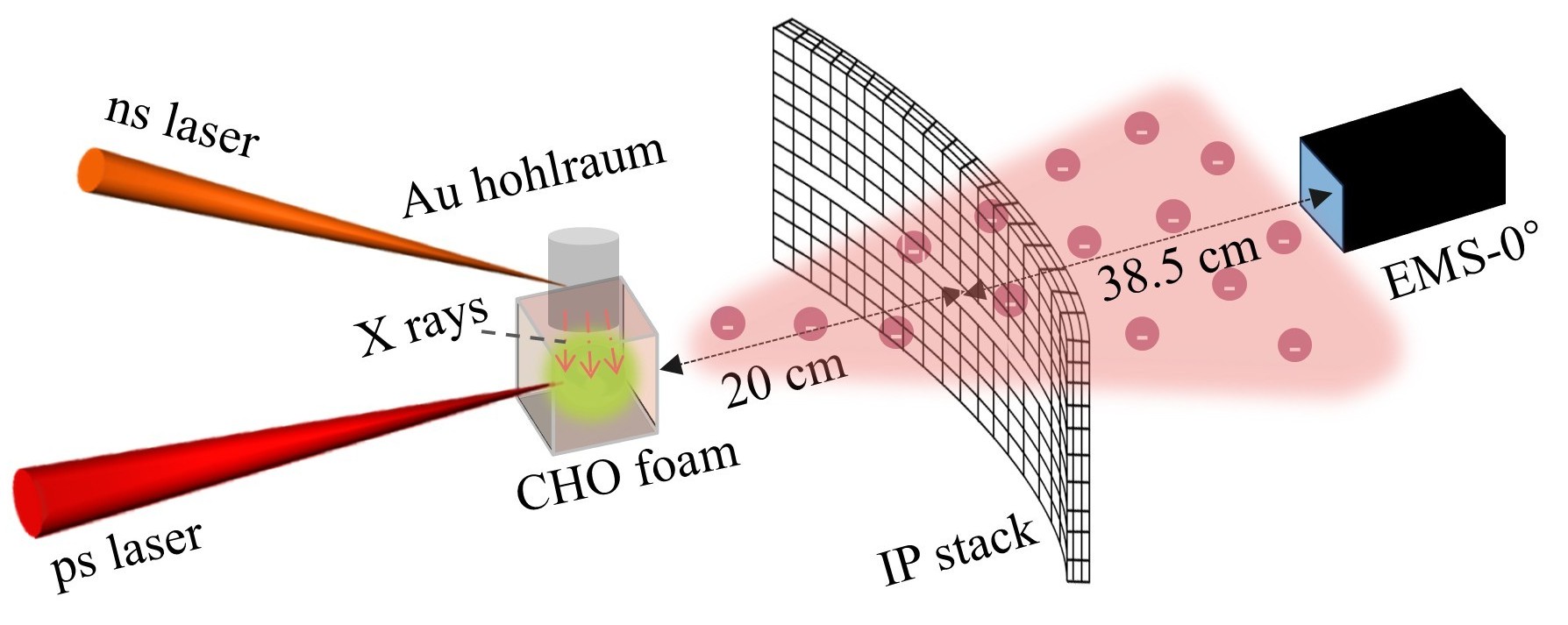}
	\caption{\label{fig:1} Schematic of the experimental setup for laser-driven electron acceleration. A nanosecond laser irradiates a gold hohlraum to generate soft X-rays, which heat the CHO foam connected to the hohlraum opening. A picosecond laser subsequently interacts with the heated CHO foam to drive electron acceleration. The resulting electron beam is diagnosed using the image plate (IP) stack and an electron magnetic spectrometer (EMS).}
\end{figure}

\section{EXPERIMENTAL SETUP}

The experiment was conducted at the XG-III laser facility, which is capable of delivering synchronized femtosecond, picosecond, and nanosecond laser pulses \cite{wuXingGuangIIILaser2020}. In this work, a coordinated dual-pulse irradiation scheme using nanosecond and picosecond lasers was employed, as illustrated in Fig.~\ref{fig:1}. A 150 J nanosecond laser irradiated a gold hohlraum to generate soft X-rays, which indirectly heated a $\mathrm{C}_9\mathrm{H}_{16}\mathrm{O}_8$ foam to a plasma state. With an initial mass density of 2 mg/cm$^3$ and a thickness of 1 mm, the plasma conditions were characterized in our previous work\cite{maPlasmaSpectroscopyHydrogenCarbonOxygen2022}. The electron temperature was about 17 eV, and the average ionization states were C$^{3.8+}$, H$^{0.98+}$, and O$^{4.5+}$. Assuming a constant mass density, we estimated the free electron density to be about 4 $\times$ 10$^{20}$ cm$^{-3}$, which is in the NCD plasma regime. 

A picosecond laser pulse with a total energy of 120 J, a focal spot full width at half maximum (FWHM) of 20 µm, and a pulse duration of 0.8 ps was subsequently focused onto the foam target to drive the DLA process for electron generation, with more than 30\% of the laser energy contained within the FWHM focal area. The time delay between the ps and ns lasers can in principle be continuously adjusted within a tens-of-nanoseconds timescale. In this work, experiments were conducted with time delays of $\tau$ = 3, 6, 9, and 15 ns, respectively. For comparison, electron beams generated by direct ps-laser irradiation of the cold foam and a Cu foil were also investigated.

A diagnostic stack consisting of a 3 mm thick cylindrical stainless steel and a Fuji Image Plate (IP)\cite{ingenitoComparativeCalibrationIP2016} was placed 20 cm from the foam target to characterize the spatial distribution of the electron beam with energy $E_k$ > 3.4 MeV. The stack had a central slit to allow electrons to pass through.  Electrons passing through the slit then traveled to an electron magnetic spectrometer (EMS) placed along the laser axis at 0° for energy spectrum measurement.

\section{RESULTS AND DISCUSSION}

\begin{figure}
	\centering
	\includegraphics[width=\linewidth]{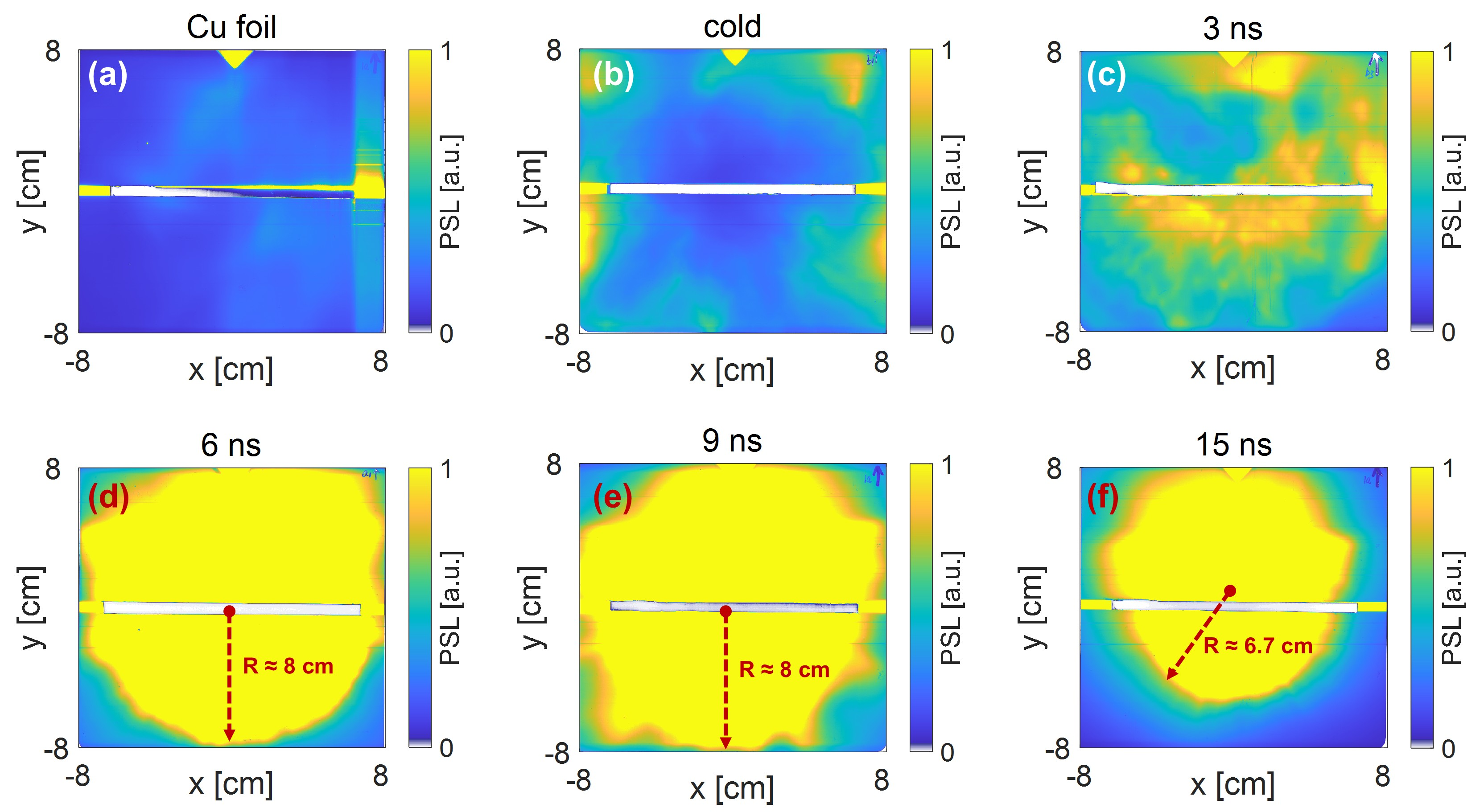}
	\caption{\label{fig:2} The spatial intensity distributions in photostimulated luminescence (PSL) of electron beams ({$E_k$ > 3.4 MeV}) driven by picosecond laser interactions with (a) Cu foil, (b) cold foam, and (c)-(f) heated foams at different delay times, respectively, based on IP measurements.}
\end{figure}

The spatial intensity distributions of the electron beams ({$E_k$ > 3.4 MeV}) measured using the IP for target cases of Cu foil, cold foam, and NCD plasmas at delays of $\tau$ = 3, 6, 9, and 15 ns are shown in Fig.~\ref{fig:2}. A unified colormap is used across all subfigures for consistent comparison. For the Cu foil target, the electron signal is extremely weak, and no distinct beam spot is observed. In the case of the cold foam target, most electrons diverge beyond the laser axis, resulting in a low-intensity distribution in the center. Once the foam is heated ($\tau$ = 3 ns), the electron beam becomes gradually collimated. At $\tau$ = 6 ns and 9 ns, the collimated beam is sufficiently intense to saturate the IP. The beam radius, which is roughly determined as the distance from center to obvious boundary as indicated in Fig.~\ref{fig:2}, is about $R$ $\approx$ 8 cm. For the $\tau$ = 15 ns case, the beam spot becomes smaller ($R$ $\approx$ 6.7 cm), and the IP remains saturated.

\begin{figure}
	\centering
	\includegraphics[width=\linewidth]{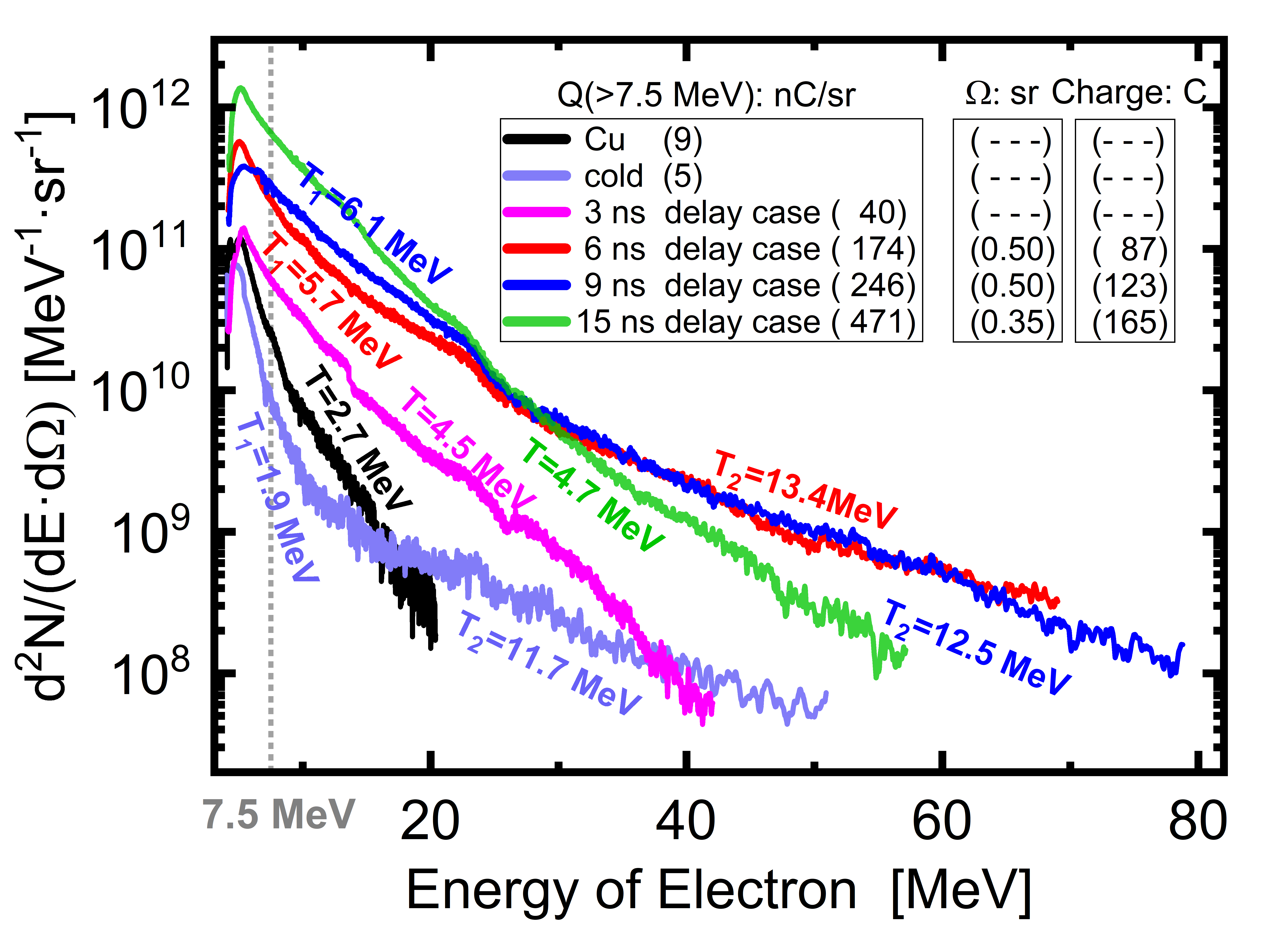}
	\caption{\label{fig:3} Differential energy spectra of electron beams along the laser propagation direction, obtained using the EMS for picosecond laser interactions with different targets. The solid angle $\Omega$ is estimated from the beam spot radius in Fig.~\ref{fig:2}. Due to the saturation of the IP board, the absolute charge values presented here are estimated upper limits.}
\end{figure}

\begin{figure}[b]
	\centering
	\includegraphics[width=7.5 cm]{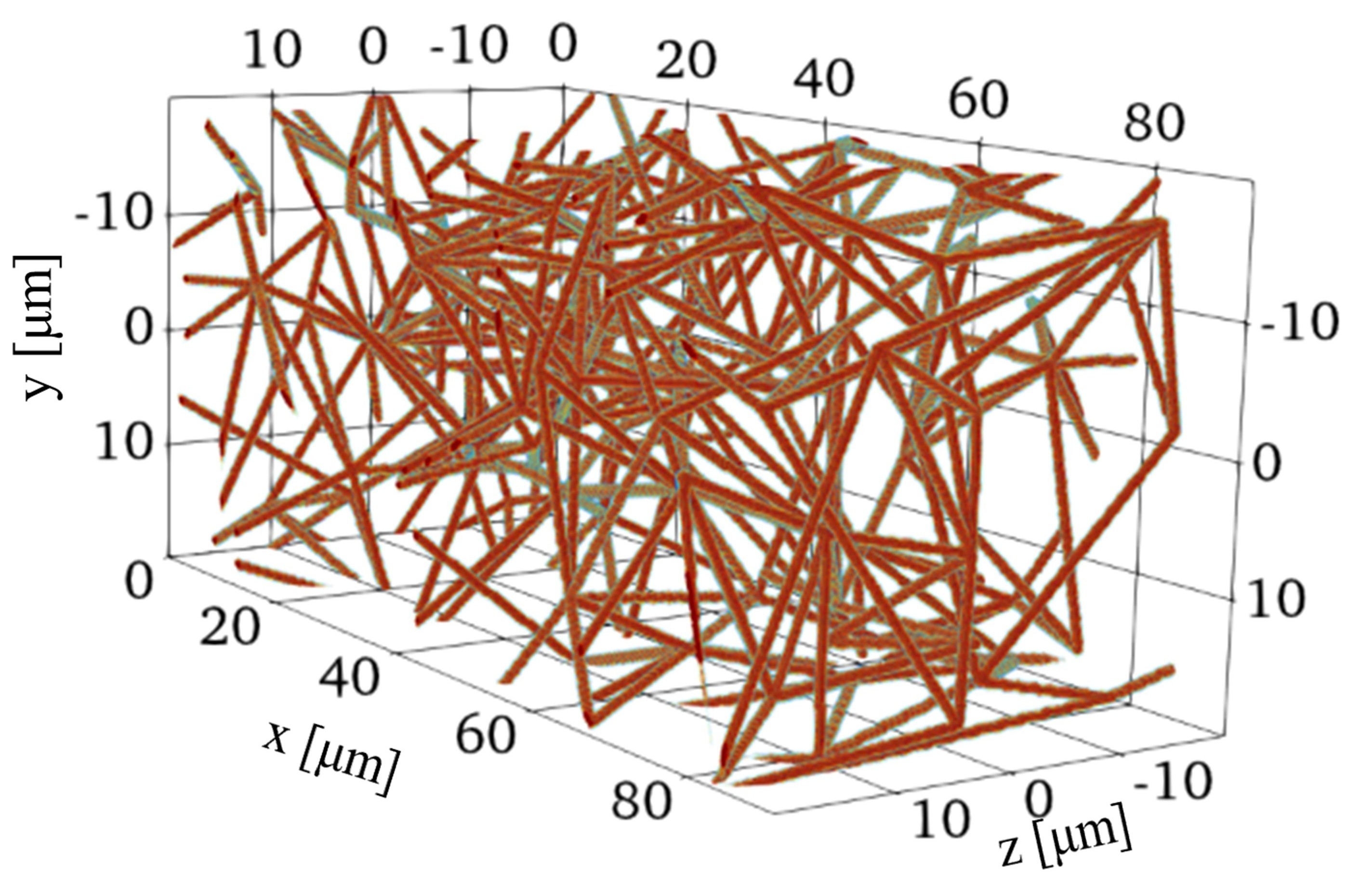}
	\caption{\label{fig:4} Initial skeleton (solid filament) density distribution of the cold foam. The electron density of the skeleton is 3.9 × 10$^{22}$ cm$^{-3}$.}
\end{figure}

\begin{figure}
	\centering
	\includegraphics[width=\linewidth]{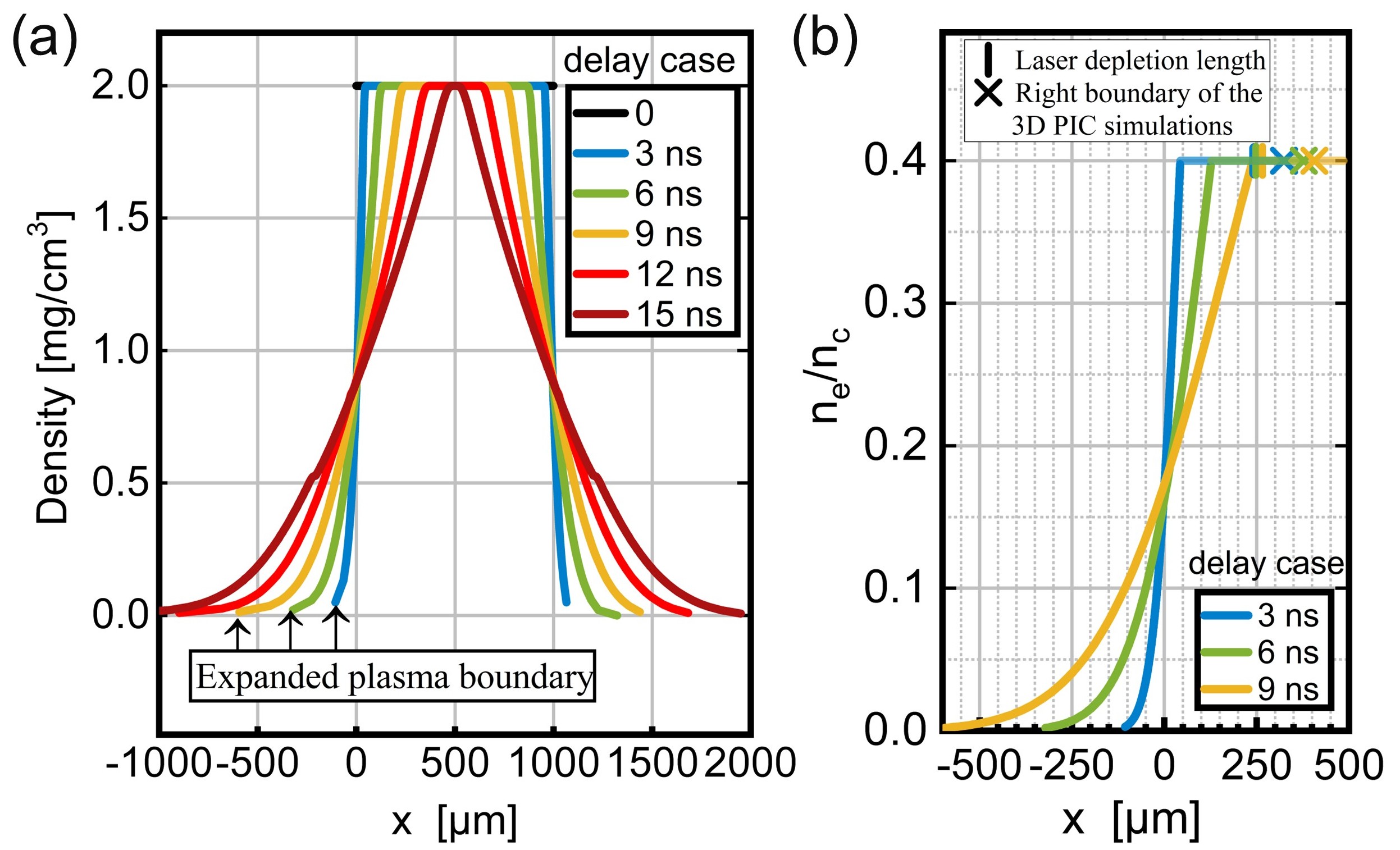}
	\caption{\label{fig:5} (a) Simulated longitudinal mass density distributions at different delay times under XG-III experimental conditions, following indirect X-ray heating by a nanosecond laser-driven gold hohlraum. The zero position corresponds to the left boundary of the foam target in its initial (unheated) state. (b) Schematic of the density profile used in the 3D PIC simulation ($n_c$ = 1.0 × 10$^{21}$ cm$^{-3}$). The picosecond laser propagates from left to right in the PIC simulation.}
\end{figure}

\begin{figure*}
	\centering
	\includegraphics[width=\linewidth]{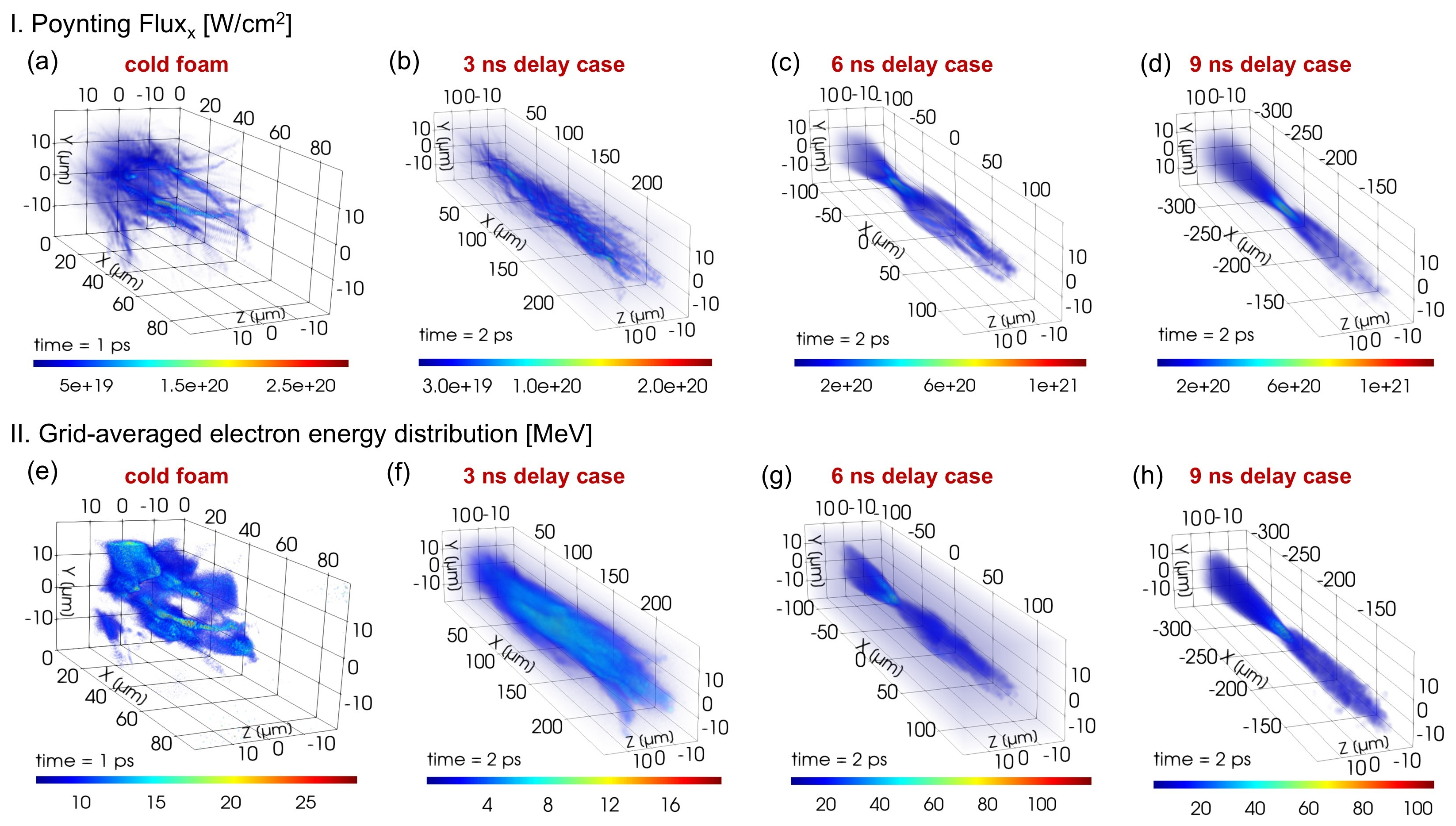}
	\caption{\label{fig:6} (I.) Simulated laser intensity (represented by the Poynting flux along the laser propagation direction) and (II.) the grid-averaged electron energy under different target conditions, respectively. The regions displayed correspond to only a part of the PIC simulation domain.}
	\vspace{2 mm}
\end{figure*}

The electron energy spectra measured along the laser axis (0°) for the Cu foil, cold foam, and NCD plasmas at delays of $\tau$ = 3, 6, 9, and 15 ns are shown in Fig.~\ref{fig:3}. For the Cu foil target, the spectrum exhibits a single-temperature distribution with an effective temperature of $T\:\approx\:$2.7 MeV and a cut-off energy of only about 20 MeV, characteristic of ponderomotive acceleration. For the cold foam target, the spectrum shows a two-temperature structure: a lower-temperature component $T_1\approx\:$1.9 MeV, similar to that of ponderomotive acceleration, and a higher-temperature component $T_2\approx\:$11.7 MeV, attributed to DLA. In the case of NCD plasmas ($\tau$ = 6 ns and 9 ns), the spectra also exhibit a two-temperature distribution, with $T_1\approx\:$6 MeV and $T_2\approx\:$13 MeV, and the cut-off energy reaches approximately 80 MeV. For the shorter delay ($\tau$ = 3 ns) and the longer delay ($\tau$ = 15 ns), the spectra become nearly single‑temperature, with effective temperatures of about 4.5 MeV and 4.7 MeV, respectively.

The total electron charge per steradian for kinetic energies above 7.5 MeV, denoted as $Q$, was obtained by integrating the energy spectra. The threshold of 7.5 MeV was chosen because it is critical for nuclear reactions in the giant dipole resonance (GDR) region. The results show a significant increase in $Q$ for the NCD plasma cases compared with the Cu foil ($Q$ $\approx$ 9 nC/sr) and cold foam targets ($Q$ $\approx$ 5 nC/sr). For the NCD plasma at $\tau$ = 15 ns, $Q$ reaches approximately 471 nC/sr. Combining this value with the beam solid angle $\Omega$ derived from the spatial distribution yields a total beam charge of 165 nC. Furthermore, for the NCD plasma cases, the charge increases with the time delay (from $\tau$ = 3 ns to 15 ns).

We performed three-dimensional numerical simulations for the acceleration processes using the PIC code EPOCH\cite{arberContemporaryParticleincellApproach2015}. The parameters of the picosecond laser in the simulations are the same as those in the experiment. The laser is linearly polarized along the $y$‑axis and propagates along the positive $x$‑axis, with a Gaussian profile in both time and space. It has a wavelength of 1.06 $\mu$m, pulse duration of 0.8 ps, spot size of 20 $\mu$m, and peak intensity of 1.4 $\times$ 10$^{19}$ W/cm$^2$.

The cold foam was modeled as a porous structure consisting of voids and a fully ionized C-H-O skeleton, as shown in Fig.~\ref{fig:4}. The porosity, defined as the ratio of the void volume to the total target volume, was 98.4\%. The skeleton had a diameter of 0.4 $\mu$m and an electron density of 3.9 × 10$^{22}$ cm$^{-3}$. These parameters ensured that the foam had the same mass density as that used in the experiments. The hydrodynamic behavior of the foam layer heated by the hohlraum radiation was simulated with a 1D hydrodynamic code\cite{guskovEquationStatePartially2023a} to establish the spatio-temporal evolution of the plasma density profile for 3D PIC modeling.

Fig.~\ref{fig:5}\textcolor{blue}{(a)} shows the simulated spatial distributions of the plasma density at different delay times after the CHO-foam was heated by soft X-ray radiation. In the PIC model, the heated plasma consists of  C$^{4+}$, H$^{+}$, and O$^{5+}$ ions and free electrons with a temperature of 17 eV and a peak density of 4 × 10$^{20}$ cm$^{-3}$ according to the experiment condition. The picosecond laser propagates from left to right along the $x$-axis. The position $x=0$ corresponds to the initial left boundary of the unheated and unexpanded foam and is also the laser focusing surface for all cases. After heating, plasma expansion occurs. The left boundary of the plasma in the PIC simulation is indicated by black arrows in Fig.~\ref{fig:5}\textcolor{blue}{(a)}, while the right boundary is indicated by crosses in Fig.~\ref{fig:5}\textcolor{blue}{(b)}. The right boundary was chosen to lie beyond the laser depletion length. The simulations were limited to the cold foam (0 ns delay, i.e., unheated) and the $\tau$ = 3 ns, 6 ns, and 9 ns delay cases, since at longer delays, the plasma expands to a larger volume, requiring significant computational resources. With these cases, the main physics can already be captured.

In the PIC simulations, each cell contained 4 macro-particles for electrons and 1 macro-particle for each ion species. The spatial resolution was 0.1 $\mu$m in the $x$‑direction. The resolutions in the $y$‑ and $z$‑directions were 0.5 $\mu$m for plasma and 0.2 $\mu$m for the cold foam case. Absorbing boundary conditions were applied for electromagnetic fields, and transparent boundary conditions for particles.
%\vspace{-4 mm} 

The laser intensity, which was represented by the Poynting flux, and the grid-averaged electron energy in the target for the cold foam case and the plasma cases with delays of 3 ns, 6 ns and 9 ns are presented in Fig.~\ref{fig:6}, respectively. Fig.~\ref{fig:6} \textcolor{blue}{(a)} shows that in the cold foam, the laser rapidly breaks into multiple divergent filaments within 1 ps. This breakup is caused by the randomly oriented skeletons. The strong refractive index contrast between the skeletons and the voids creates a disordered array of discrete scatterers, preventing the formation of a single self‑guided channel. Consequently, the electron energy distribution, as shown in Fig.~\ref{fig:6} \textcolor{blue}{(e)}, is widely spread over space as well. This is consistent with the experimentally observed large divergence in Fig.~\ref{fig:2} \textcolor{blue}{(b)}.
%\vspace{-4 mm} 

After heating, the foam target becomes homogenized, allowing the laser to propagate in a better-collimated channel via relativistic self-focusing \cite{litvak1970finite,maxSelfModulationSelfFocusingElectromagnetic1974,siegristSelffocusingPlasmaDue1976,sunSelffocusingShortIntense1987,sprangleRelativisticSelfFocusingShortPulse1987,esareyPhysicsLaserdrivenPlasmabased2009}, as illustrated in Figs.~\ref{fig:6}\textcolor{blue}{(b)–(d)}. The plasma density gradient varies with the time delay, forming density ramps of different steepness that jointly determine laser transport \cite{sadighi-bonabiImprovingRelativisticSelffocusing2009} and the resultant electron beam characteristics \cite{rosmejAdvancedPlasmaTarget2025}.

For relativistic self-focusing of laser in underdense plasma, the characteristic self-focusing length reads \cite{zengSelftruncatedIonizationInjection2014,zouEnhancedTargetNormal2014}:
\begin{equation}
	L_{\mathrm{sf}}\approx \frac{Z_\mathrm{R}}{\sqrt{P/P_c-1}}
	=\frac{\pi w_0^2}{\lambda_0\sqrt{P/P_c-1}},
\end{equation}
where \(Z_\mathrm{R}\) is the Rayleigh length, \(w_0\) is the initial laser spot radius, and \(P_c\) is the critical power for relativistic self-focusing, given by\cite{sprangleRelativisticSelfFocusingShortPulse1987}
\begin{equation}
	P_c \approx 17\frac{n_c}{n_e}\,\mathrm{GW},
\end{equation}
where \(n_c\) is the critical density and \(n_e\) is the local electron density. Here, \(n_c\) = 1.0 × 10$^{21}$ cm$^{-3}$.

According to equations (1) and (2), as the laser propagates into the plasma, the local density $n_e$ increases, and the local self-focusing length $L_{\mathrm{sf}}$ decreases accordingly. The density gradient thus governs the shrink rate of $L_{\mathrm{sf}}$ along the beam propagation path. For the laser parameters used in our experiment, $L_{\mathrm{sf}}$ falls from 337 $\mu$m to 22 $\mu$m when plasma density increases from the simulation boundary $n_e$ = 0.002 $n_c$ to the peak value $n_e$ = 4 $\times$ 10$^{20}$ cm$^{-3}$ = 0.4 $n_c$.

For the steepest ramp case ($\tau$ = 3 ns),  the electron density jumps from 0.002 $n_c$ to 0.4 $n_c$ within a short ramp length $L_{\mathrm{ramp}}$ $\approx$ 149 $\mu$m. Correspondingly, $L_\mathrm{sf}$ plummets sharply across the narrow spatial range. Such abrupt beam compression amplifies transverse modulations and filamentation instabilities driven by nonlinear plasma effects \cite{lezhninParallelLaserBeam2024,huangLargeangleStimulatedRaman2025}. Simulations confirm that the laser undergoes rapid self-focusing after propagating roughly 100 $\mu$m, then quickly breaks into discrete filaments.

In contrast, for mild density ramps cases of  $\tau$ = 6 ns and $\tau$ = 9 ns, the ramp lengths is longer as $L_\mathrm{ramp}$ $\approx$ 448 $\mu$m and $L_\mathrm{ramp}$ $\approx$ 833 $\mu$m, respectively. The gradual variation of $L_\mathrm{sf}$ over longer distance enables the laser beam to maintain a quasi-equilibrium profile during propagation, accompanied by a slowly growing focusing force. This stabilizes laser transport over extended distances. The simulation shows that the effective self-focusing distance stretches to 200--300 $\mu$m, forming a single, collimated propagation channel. The corresponding electron energy distribution [Figs.~\ref{fig:6}\textcolor{blue}{(g)–(h)}] concentrates tightly on the propagation axis, consistent with the highly collimated electron beams observed experimentally.

\begin{figure}
	\centering
	\includegraphics[width=8.1 cm]{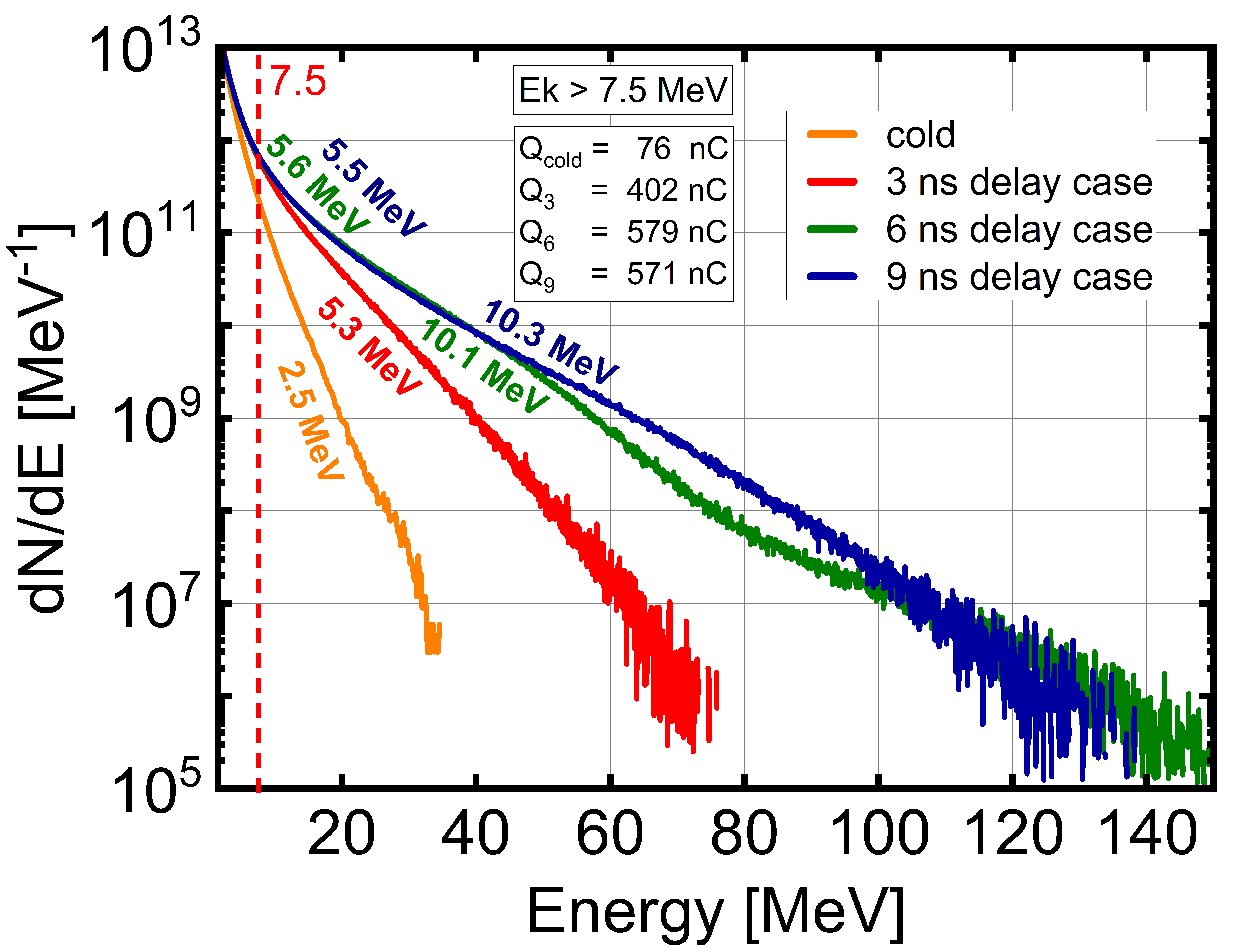}
	\caption{\label{fig:7} Simulated electron energy spectra of in-target and escaped electrons with momentum $p_x > 0$.}
\end{figure}

All forward‑propagating electrons (with momentum $p_x > 0$) inside the target and those exiting the simulation box were collected to obtain the energy spectra and to evaluate the electron beam temperature and charge. The results are presented in Fig.~\ref{fig:7}. The effective temperature and cut-off energy of the electron beams are significantly higher for the $\tau$ = 6 ns and 9 ns plasma cases in comparison with the cold foam or $\tau$ = 3 ns plasma case. This improvement is attributed to the longer propagation distance and better‑guided laser channel, which enhance DLA. The simulated temperatures of $T$ > 10 MeV ($\tau$ = 6 ns, 9 ns), and $T$ $\approx$ 5.3 MeV ($\tau$ = 3 ns) are on the same order of magnitude as the experimental measurements ($T_2\approx\:$13 MeV for $\tau$ = 6--9 ns and $T$ $\approx$ 4.5 MeV for $\tau$ = 3 ns). The charge of electrons with energies above 7.5 MeV was also analyzed. The results show  that the total charge generally increases with the time delay, but with nearly identical values at $\tau$ = 6 ns and 9 ns. The total charge magnitude is on the order of hundreds of nC. The magnitude agrees with the experimental measurements.

\begin{figure}
	\centering
	\includegraphics[width=8.4 cm]{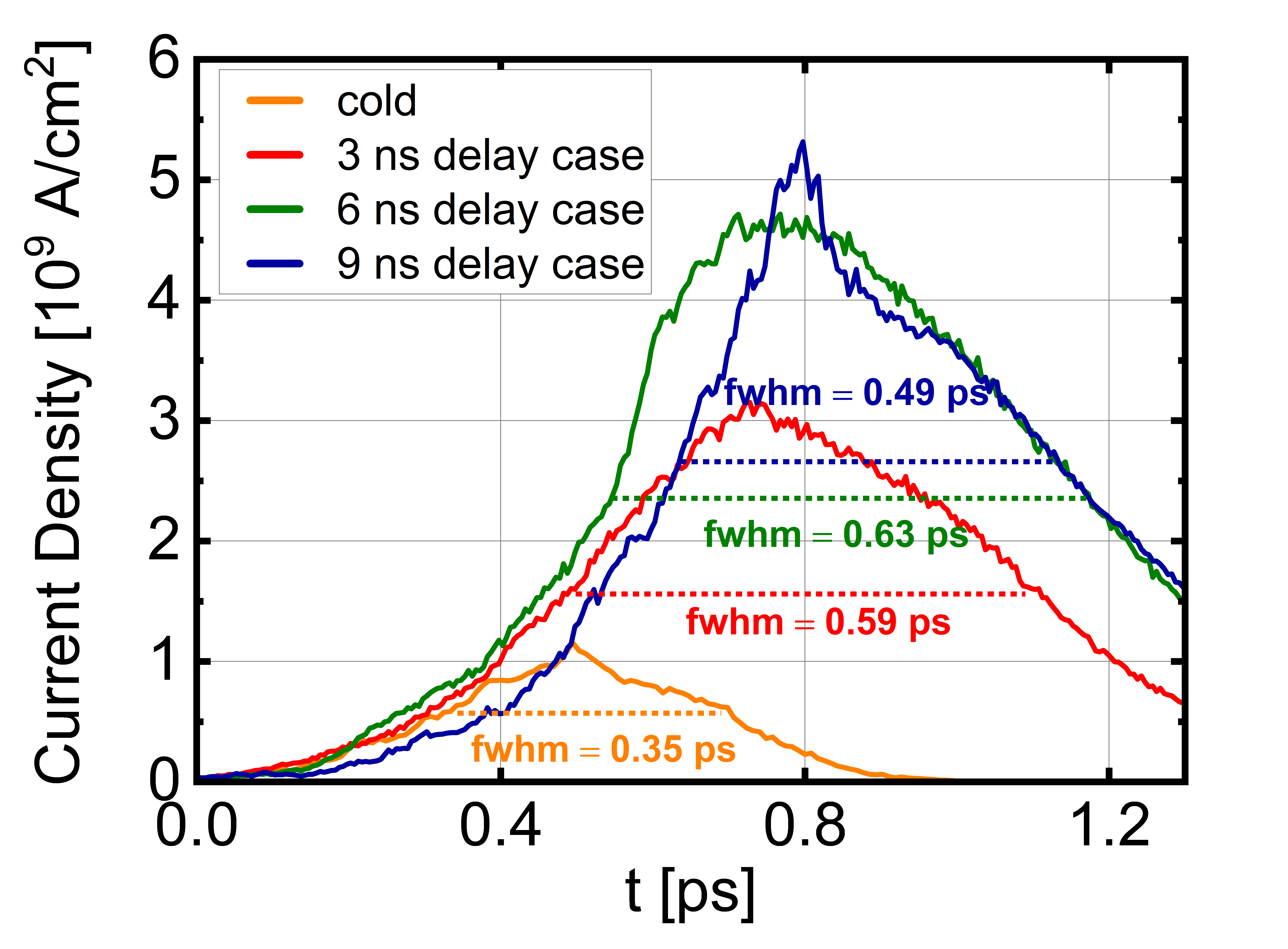}
	\caption{\label{fig:8} Current density of the electron beam ($E_k$ > 7.5 MeV) recorded at the right diagnostic boundary. \textit{t} = 0 corresponds to the moment when the electron beam first arrives at the right boundary of the simulation box.}
\end{figure}

The pulse duration and intensity of the electron beam with $E_k$ > 7.5 MeV were additionally obtained through counting the number of electrons hitting the right boundary of the simulation box at different times. Fig.~\ref{fig:8} shows the current density of the electron beam, where $t$ = 0 corresponds to the moment when the first electron beam arrives at the right boundary. The duration, which is defined as the FWHM of the pulse, is on the order of hundreds of fs. Similarly to the electron charge variation trend, the current density peak is highest for the $\tau$ = 6 ns and 9 ns cases, and reaches 5 $\times$ 10$^9$ A/cm$^2$.

%\vspace{-0.5 mm} 

\section{CONCLUSION}

In summary, we demonstrate tunable generation of high-charge relativistic electron beams via direct laser acceleration (DLA) in near-critical-density (NCD) plasmas produced by indirectly heating porous foam targets with hohlraum soft X-ray. By adjusting the delay $\tau$ between the nanosecond and picosecond lasers, we manipulate the plasma density profile and, consequently, the electron beam parameters. 

Experiments show that the preheated NCD plasma yields a one to two order-of-magnitude enhancement in beam charge compared with both solid foil and cold foam targets, alongside a substantial increase in effective temperature and cutoff energy. This performance improvement reflects a transition from dominant pondermotive acceleration to efficient DLA enabled by homogenized plasma conditions. At delays of 6–9 ns, the electron beam reaches an effective temperature of ~13 MeV and a cutoff energy of ~80 MeV. As the delay extends from 3 ns to 15 ns, the charge per steradian for electrons above 7.5 MeV rises from tens to 471 nC/sr, while the effective temperature first increases and then declines, revealing an inherent trade-off between total beam charge and spectral hardness.
Three-dimensional particle-in-cell simulations consistently reproduce the experimental trends, and uncover that the interplay between the intrinsic foam structure and the time-evolving density gradient governs laser relativistic self-focusing and channel formation, which ultimately determines the DLA efficiency and final beam properties. This work identifies the critical physical mechanism linking target structure, density ramp and DLA performance, and provides a reliable experimental pathway to customize electron beam temperature, charge and divergence. The tunable high-charge relativistic electron source offers broad application prospects in high-energy-density physics and photonuclear reaction research.

\section*{ACKNOWLEDGMENTS}
This study was financially supported by the National Key R\&D Program of China under Grant No.~2022YFA1603300, the Chinese Science Challenge Project under Grant No.~TZ2025012, the National Natural Science Foundation of China under Grant Nos.~12422512, 12595365, U2541245, 12120101005, 12405238, 12175174, 12325406, and 92261201, the Postdoctoral Fellowship Program of CPSF under Grant No.~GZC20241372, the Postdoctoral Research Funding Project of Shaanxi Province under Grant No.~2024BSHYDZZ014, the ENN's Hydrogen Boron Fusion Research Fund under Grant No.~2025ENNHB01-013, the China Postdoctoral Science Foundation under Grant No.~2024M762569, the Shaanxi Fundamental Science Research Project for Mathematics and Physics under Grant No.~25JSQ025, the Fundamental Research Funds for the Central Universities under Grant No.~xzy012025080, the Shaanxi Province Key R\&D Program under Grant No.~2024PT-ZCK-83, and the Innovative Scientific Program of CNNC.

\section*{AUTHOR DECLARATIONS}

\subsection*{Conflict of Interest}
The authors have no conflicts to disclose.
\subsection*{Author Contributions}
\noindent
\textbf{Ziyao Wang}: Data curation (equal); Formal analysis (equal); Software (equal); Visualization (lead); Writing -- original draft (lead); Writing -- review \& editing (equal). \textbf{Jieru Ren}: Conceptualization (equal); Formal analysis (equal); Funding acquisition (equal); Project administration (equal); Writing -- review \& editing (equal). \textbf{Zhigang Deng}: Conceptualization (equal); Formal analysis (equal); Data curation (equal); Resources (equal). \textbf{Wenqing Wei}: Formal analysis (equal); Methodology (equal); Investigation (equal); Project administration (equal); Writing -- review \& editing (equal). \textbf{Wei Qi}: Investigation (equal); Methodology (equal). \textbf{Olga N. Rosmej}: Validation (equal); Writing -- review \& editing (equal). \textbf{Nikolay E. Andreev}: Validation (equal); Software (equal). \textbf{Sergey Yu. Gus'kov}: Software (equal). \textbf{Rafael Yakhin}: Software (equal). \textbf{Yifang Gao}: Investigation (equal). \textbf{Bubo Ma}: Investigation (equal). \textbf{Mingzhe Yang}: Investigation (equal). \textbf{Shizheng Zhang}: Investigation (equal). \textbf{Xuyang Luo}: Investigation (equal). \textbf{Dieter H.H. Hoffmann}: Writing -- review \& editing (equal). \textbf{Peng Zhou}: Investigation (equal). \textbf{Ke Jiang}: Software (equal). \textbf{Taiwu Huang}: Software (equal). \textbf{Bo Cui}: Methodology (equal). \textbf{Weiwu Wang}: Methodology (equal). \textbf{Shaoyi Wang}: Methodology (equal). \textbf{Quanping Fan}: Methodology (equal). \textbf{Zhurong Cao}: Methodology (equal). \textbf{Sixin Wu}: Investigation (equal). \textbf{Yue Yang}: Investigation (equal). \textbf{Leifeng Cao}: Methodology (equal). \textbf{Yuqiu Gu}: Resources (equal). \textbf{Yuchi Wu}: Resources (equal). \textbf{Weimin Zhou}: Resources (equal). \textbf{Zongqing Zhao}: Resources (equal). \textbf{Yongtao Zhao}: Funding acquisition (equal); Supervision (lead); Writing -- review \& editing (equal).
\section*{DATA AVAILABILITY}

The data that support the findings of this study are available from the corresponding author upon reasonable request.

\section*{REFERENCES}
\bibliography{FOAM_ELECTRON_REF}% Produces the bibliography via BibTeX.

@article{arberContemporaryParticleincellApproach2015,
  title = {Contemporary Particle-in-Cell Approach to Laser-Plasma Modelling},
  author = {Arber, T D and Bennett, K and Brady, C S and {Lawrence-Douglas}, A and Ramsay, M G and Sircombe, N J and Gillies, P and Evans, R G and Schmitz, H and Bell, A R and Ridgers, C P},
  year = {2015},
  month = nov,
  journal = {Plasma Physics and Controlled Fusion},
  volume = {57},
  number = {11},
  pages = {113001},
  issn = {0741-3335, 1361-6587},
  doi = {10.1088/0741-3335/57/11/113001},
  urldate = {2025-03-19},
  langid = {english}
}

@article{babjakDirectLaserAcceleration2024b,
  title = {Direct Laser Acceleration in Varying Plasma Density Profiles},
  author = {Babjak, R and Martinez, B and Krus, M and Vranic, M},
  year = {2024},
  month = sep,
  journal = {New Journal of Physics},
  volume = {26},
  number = {9},
  pages = {093002},
  issn = {1367-2630},
  doi = {10.1088/1367-2630/ad7280},
  urldate = {2025-07-06}
}

@article{chenDensityTemperatureCharacterization2016,
  title = {Density and Temperature Characterization of Long-Scale Length, near-Critical Density Controlled Plasma Produced from Ultra-Low Density Plastic Foam},
  author = {Chen, S. N. and Iwawaki, T. and Morita, K. and Antici, P. and Baton, S. D. and Filippi, F. and Habara, H. and Nakatsutsumi, M. and Nicola{\"i}, P. and Nazarov, W. and Rousseaux, C. and Starodubstev, M. and Tanaka, K. A. and Fuchs, J.},
  year = {2016},
  month = feb,
  journal = {Scientific Reports},
  volume = {6},
  number = {1},
  pages = {21495},
  issn = {2045-2322},
  doi = {10.1038/srep21495},
  urldate = {2025-07-06},
  langid = {english}
}

@article{cikhardtCharacterizationBrightBetatron2024,
  title = {Characterization of Bright Betatron Radiation Generated by Direct Laser Acceleration of Electrons in Plasma of near Critical Density},
  author = {Cikhardt, J. and Gyrdymov, M. and Z{\"a}hter, S. and Tavana, P. and G{\"u}nther, M. M. and Bukharskii, N. and Borisenko, N. and Jacoby, J. and Shen, X. F. and Pukhov, A. and Andreev, N. E. and Rosmej, O. N.},
  year = {2024},
  month = mar,
  journal = {Matter and Radiation at Extremes},
  volume = {9},
  number = {2},
  pages = {027201},
  issn = {2468-2047, 2468-080X},
  doi = {10.1063/5.0181119},
  urldate = {2025-03-17},
  langid = {english}
}

@article{cohenUndepletedDirectLaser2024a,
  title = {Undepleted Direct Laser Acceleration},
  author = {Cohen, Itamar and Meir, Talia and Tangtartharakul, Kavin and Perelmutter, Lior and Elkind, Michal and Gershuni, Yonatan and Levanon, Assaf and Arefiev, Alexey V. and Pomerantz, Ishay},
  year = {2024},
  month = jan,
  journal = {Science Advances},
  volume = {10},
  number = {2},
  pages = {eadk1947},
  issn = {2375-2548},
  doi = {10.1126/sciadv.adk1947},
  urldate = {2025-07-06},
  langid = {english}
}

@article{courtoisHighresolutionMultiMeVXray2011,
  title = {High-Resolution Multi-{{MeV}} x-Ray Radiography Using Relativistic Laser-Solid Interaction},
  author = {Courtois, C. and Edwards, R. and Compant La Fontaine, A. and Aedy, C. and Barbotin, M. and Bazzoli, S. and Biddle, L. and Brebion, D. and Bourgade, J. L. and Drew, D. and Fox, M. and Gardner, M. and Gazave, J. and Lagrange, J. M. and Landoas, O. and Le Dain, L. and Lefebvre, E. and Mastrosimone, D. and Pichoff, N. and Pien, G. and Ramsay, M. and Simons, A. and Sircombe, N. and Stoeckl, C. and Thorp, K.},
  year = {2011},
  month = feb,
  journal = {Physics of Plasmas},
  volume = {18},
  number = {2},
  pages = {023101},
  issn = {1070-664X, 1089-7674},
  doi = {10.1063/1.3551738},
  urldate = {2025-07-05},
  langid = {english}
}

@article{dunneEvaluationFoamBuffer1995,
  title = {Evaluation of a {{Foam Buffer Target Design}} for {{Spatially Uniform Ablation}} of {{Laser-Irradiated Plasmas}}},
  author = {Dunne, M. and Borghesi, M. and Iwase, A. and Jones, M. W. and Taylor, R. and Willi, O. and Gibson, R. and Goldman, S. R. and Mack, J. and Watt, R. G.},
  year = {1995},
  month = nov,
  journal = {Physical Review Letters},
  volume = {75},
  number = {21},
  pages = {3858--3861},
  issn = {0031-9007, 1079-7114},
  doi = {10.1103/PhysRevLett.75.3858},
  urldate = {2025-07-06},
  copyright = {http://link.aps.org/licenses/aps-default-license},
  langid = {english}
}

@article{fengHighefficiencyNeutronSource2020,
  title = {High-Efficiency Neutron Source Generation from Photonuclear Reactions Driven by Laser Plasma Accelerator},
  author = {Feng, Jie and Fu, Changbo and Li, Yifei and Zhang, Xiaopeng and Wang, Jinguang and Li, Dazhang and Zhu, Changqing and Tan, Junhao and Mirzaie, Mohammad and Zhang, Zhe and Chen, Liming},
  year = {2020},
  month = aug,
  journal = {High Energy Density Physics},
  volume = {36},
  pages = {100753},
  issn = {15741818},
  doi = {10.1016/j.hedp.2020.100753},
  urldate = {2025-03-20},
  langid = {english}
}

@article{flaccoLaserDrivenFLASH2025,
  title = {Laser Driven {{FLASH}} Radiobiology Using a High Dose and Ultra High Dose Rate Single Pulse Proton Source},
  author = {Flacco, A. and Bayart, E. and Romagnani, L. and Cavallone, M. and De Marzi, L. and Fouillade, C. and Giaccaglia, C. and Heinrich, S. and {Lamarre-Jouenne}, I. and Monzac, J. and Parodi, K. and Patriarca, A. and R{\"o}sch, T. and Schreiber, J. and Tischendorf, L.},
  year = {2025},
  month = may,
  journal = {Scientific Reports},
  volume = {15},
  number = {1},
  pages = {16511},
  issn = {2045-2322},
  doi = {10.1038/s41598-025-01105-z},
  urldate = {2025-07-05},
  langid = {english}
}

@article{gahnMultiMeVElectronBeam1999a,
  title = {Multi-{{MeV Electron Beam Generation}} by {{Direct Laser Acceleration}} in {{High-Density Plasma Channels}}},
  author = {Gahn, C. and Tsakiris, G. D. and Pukhov, A. and {Meyer-ter-Vehn}, J. and Pretzler, G. and Thirolf, P. and Habs, D. and Witte, K. J.},
  year = {1999},
  month = dec,
  journal = {Physical Review Letters},
  volume = {83},
  number = {23},
  pages = {4772--4775},
  issn = {0031-9007, 1079-7114},
  doi = {10.1103/PhysRevLett.83.4772},
  urldate = {2025-07-05},
  copyright = {http://link.aps.org/licenses/aps-default-license},
  langid = {english}
}

@article{guntherForwardlookingInsightsLasergenerated2022,
  title = {Forward-Looking Insights in Laser-Generated Ultra-Intense {$\gamma$}-Ray and Neutron Sources for Nuclear Application and Science},
  author = {G{\"u}nther, M. M. and Rosmej, O. N. and Tavana, P. and Gyrdymov, M. and Skobliakov, A. and Kantsyrev, A. and Z{\"a}hter, S. and Borisenko, N. G. and Pukhov, A. and Andreev, N. E.},
  year = {2022},
  month = jan,
  journal = {Nature Communications},
  volume = {13},
  number = {1},
  pages = {170},
  issn = {2041-1723},
  doi = {10.1038/s41467-021-27694-7},
  urldate = {2025-03-17},
  langid = {english}
}

@article{huangRelativisticLaserHosing2017a,
  title = {Relativistic Laser Hosing Instability Suppression and Electron Acceleration in a Preformed Plasma Channel},
  author = {Huang, T. W. and Zhou, C. T. and Zhang, H. and Wu, S. Z. and Qiao, B. and He, X. T. and Ruan, S. C.},
  year = {2017},
  month = apr,
  journal = {Physical Review E},
  volume = {95},
  number = {4},
  pages = {043207},
  issn = {2470-0045, 2470-0053},
  doi = {10.1103/PhysRevE.95.043207},
  urldate = {2025-07-06},
  copyright = {http://link.aps.org/licenses/aps-default-license},
  langid = {english}
}

@article{husseinOptimisationDirectLaser2021a,
  title = {Towards the Optimisation of Direct Laser Acceleration},
  author = {Hussein, A E and Arefiev, A V and Batson, T and Chen, H and Craxton, R S and Davies, A S and Froula, D H and Gong, Z and Haberberger, D and Ma, Y and Nilson, P M and Theobald, W and Wang, T and Weichman, K and Williams, G J and Willingale, L},
  year = {2021},
  month = feb,
  journal = {New Journal of Physics},
  volume = {23},
  number = {2},
  pages = {023031},
  issn = {1367-2630},
  doi = {10.1088/1367-2630/abdf9a},
  urldate = {2025-07-06}
}

@article{ingenitoComparativeCalibrationIP2016,
  title = {Comparative Calibration of {{IP}} Scanning Equipment},
  author = {Ingenito, F. and Andreoli, P. and Batani, D. and Boutoux, G. and Cipriani, M. and Consoli, F. and Cristofari, G. and Curcio, A. and Angelis, R. De and Giorgio, G. Di and Ducret, J. and {Forestier-Colleoni}, P. and Hulin, S. and Jakubowska, K. and Rabhi, N.},
  year = {2016},
  month = may,
  journal = {Journal of Instrumentation},
  volume = {11},
  number = {05},
  pages = {C05012-C05012},
  issn = {1748-0221},
  doi = {10.1088/1748-0221/11/05/C05012},
  urldate = {2025-07-06},
  copyright = {http://iopscience.iop.org/info/page/text-and-data-mining}
}

@article{kimStableMultiGeVElectron2017a,
  title = {Stable Multi-{{GeV}} Electron Accelerator Driven by Waveform-Controlled {{PW}} Laser Pulses},
  author = {Kim, Hyung Taek and Pathak, V. B. and Hong Pae, Ki and Lifschitz, A. and Sylla, F. and Shin, Jung Hun and Hojbota, C. and Lee, Seong Ku and Sung, Jae Hee and Lee, Hwang Woon and Guillaume, E. and Thaury, C. and Nakajima, Kazuhisa and Vieira, J. and Silva, L. O. and Malka, V. and Nam, Chang Hee},
  year = {2017},
  month = aug,
  journal = {Scientific Reports},
  volume = {7},
  number = {1},
  pages = {10203},
  issn = {2045-2322},
  doi = {10.1038/s41598-017-09267-1},
  urldate = {2025-07-05},
  langid = {english}
}

@article{kneipBrightSpatiallyCoherent2010,
  title = {Bright Spatially Coherent Synchrotron {{X-rays}} from a Table-Top Source},
  author = {Kneip, S. and McGuffey, C. and Martins, J. L. and Martins, S. F. and Bellei, C. and Chvykov, V. and Dollar, F. and Fonseca, R. and Huntington, C. and Kalintchenko, G. and Maksimchuk, A. and Mangles, S. P. D. and Matsuoka, T. and Nagel, S. R. and Palmer, C. A. J. and Schreiber, J. and Phuoc, K. Ta and Thomas, A. G. R. and Yanovsky, V. and Silva, L. O. and Krushelnick, K. and Najmudin, Z.},
  year = {2010},
  month = dec,
  journal = {Nature Physics},
  volume = {6},
  number = {12},
  pages = {980--983},
  issn = {1745-2473, 1745-2481},
  doi = {10.1038/nphys1789},
  urldate = {2025-07-05},
  langid = {english}
}

@article{kneipObservationSynchrotronRadiation2008a,
  title = {Observation of {{Synchrotron Radiation}} from {{Electrons Accelerated}} in a {{Petawatt-Laser-Generated Plasma Cavity}}},
  author = {Kneip, S. and Nagel, S. R. and Bellei, C. and Bourgeois, N. and Dangor, A. E. and Gopal, A. and Heathcote, R. and Mangles, S. P. D. and Marqu{\`e}s, J. R. and Maksimchuk, A. and Nilson, P. M. and Phuoc, K. Ta and Reed, S. and Tzoufras, M. and Tsung, F. S. and Willingale, L. and Mori, W. B. and Rousse, A. and Krushelnick, K. and Najmudin, Z.},
  year = {2008},
  month = mar,
  journal = {Physical Review Letters},
  volume = {100},
  number = {10},
  pages = {105006},
  issn = {0031-9007, 1079-7114},
  doi = {10.1103/PhysRevLett.100.105006},
  urldate = {2025-07-06},
  copyright = {http://link.aps.org/licenses/aps-default-license},
  langid = {english}
}

@article{leemansMultiGeVElectronBeams2014a,
  title = {Multi-{{GeV Electron Beams}} from {{Capillary-Discharge-Guided Subpetawatt Laser Pulses}} in the {{Self-Trapping Regime}}},
  author = {Leemans, W. P. and Gonsalves, A. J. and Mao, H.-S. and Nakamura, K. and Benedetti, C. and Schroeder, C. B. and T{\'o}th, {\relax Cs}. and Daniels, J. and Mittelberger, D. E. and Bulanov, S. S. and Vay, J.-L. and Geddes, C. G. R. and Esarey, E.},
  year = {2014},
  month = dec,
  journal = {Physical Review Letters},
  volume = {113},
  number = {24},
  pages = {245002},
  issn = {0031-9007, 1079-7114},
  doi = {10.1103/PhysRevLett.113.245002},
  urldate = {2025-07-05},
  copyright = {http://link.aps.org/licenses/aps-default-license},
  langid = {english}
}

@article{maPhotonuclearProductionMedical2019,
  title = {Photonuclear Production of Medical Isotopes 62,{{64Cu}} Using Intense Laser-Plasma Electron Source},
  author = {Ma, ZhiGuo and Lan, HaoYang and Liu, WeiYuan and Wu, ShaoDong and Xu, Yi and Zhu, ZhiChao and Luo, Wen},
  year = {2019},
  month = nov,
  journal = {Matter and Radiation at Extremes},
  volume = {4},
  number = {6},
  pages = {064401},
  issn = {2468-2047, 2468-080X},
  doi = {10.1063/1.5100925},
  urldate = {2025-03-17},
  langid = {english}
}

@article{maPlasmaSpectroscopyHydrogenCarbonOxygen2022,
  title = {Plasma {{Spectroscopy}} on {{Hydrogen-Carbon-Oxygen Foam Targets Driven}} by {{Laser-Generated Hohlraum Radiation}}},
  author = {Ma, Bubo and Ren, Jieru and Wang, Shaoyi and Wang, Xing and Yin, Shuai and Feng, Jianhua and Wei, Wenqing and Xu, Xing and Chen, Benzheng and Zhang, Shisheng and Xu, Zhongfeng and Hu, Zhongmin and Li, Fangfang and Xu, Hao and Li, Taotao and Li, Yutian and Wang, Yingying and Liu, Lirong and Liu, Wei and Fan, Quanping and Chen, Yong and Deng, Zhigang and Qi, Wei and Cui, Bo and Zhou, Weimin and Zhao, Zongqing and Cao, Zhurong and Gu, Yuqiu and Cao, Leifeng and Cheng, Rui and Xue, Quanxi and Hoffmann, Dieter H. H. and Zhao, Yongtao and Batani, Dimitri},
  year = {2022},
  journal = {Laser and Particle Beams},
  volume = {2022},
  pages = {e15},
  issn = {0263-0346, 1469-803X},
  doi = {10.1155/2022/3049749},
  urldate = {2025-03-20},
  langid = {english}
}

@article{maUltrahighchargeElectronBeams2018,
  title = {Ultrahigh-Charge Electron Beams from Laser-Irradiated Solid Surface},
  author = {Ma, Yong and Zhao, Jiarui and Li, Yifei and Li, Dazhang and Chen, Liming and Liu, Jianxun and Dann, Stephen J. D. and Ma, Yanyun and Yang, Xiaohu and Ge, Zheyi and Sheng, Zhengming and Zhang, Jie},
  year = {2018},
  month = jul,
  journal = {Proceedings of the National Academy of Sciences},
  volume = {115},
  number = {27},
  pages = {6980--6985},
  issn = {0027-8424, 1091-6490},
  doi = {10.1073/pnas.1800668115},
  urldate = {2025-03-07},
  langid = {english}
}

@article{paganoSourceSizeRays2024,
  title = {Source Size of x Rays from Self-Modulated Laser Wakefield Accelerators},
  author = {Pagano, I. M. and Lemos, N. and King, P. M. and Rusby, D. and Sinclair, M. and Aghedo, A. and Khan, S. and Downer, M. C. and Joshi, C. and Albert, F.},
  year = {2024},
  month = jul,
  journal = {Physics of Plasmas},
  volume = {31},
  number = {7},
  pages = {073110},
  issn = {1070-664X, 1089-7674},
  doi = {10.1063/5.0191435},
  urldate = {2025-03-20},
  langid = {english}
}

@article{pomerantzUltrashortPulsedNeutron2014,
  title = {Ultrashort {{Pulsed Neutron Source}}},
  author = {Pomerantz, I. and McCary, E. and Meadows, A. R. and Arefiev, A. and Bernstein, A. C. and Chester, C. and Cortez, J. and Donovan, M. E. and Dyer, G. and Gaul, E. W. and Hamilton, D. and Kuk, D. and Lestrade, A. C. and Wang, C. and Ditmire, T. and Hegelich, B. M.},
  year = {2014},
  month = oct,
  journal = {Physical Review Letters},
  volume = {113},
  number = {18},
  pages = {184801},
  issn = {0031-9007, 1079-7114},
  doi = {10.1103/PhysRevLett.113.184801},
  urldate = {2025-07-05},
  copyright = {http://link.aps.org/licenses/aps-default-license},
  langid = {english}
}

@article{pugachevAccelerationElectronsAction2016a,
  title = {Acceleration of Electrons under the Action of Petawatt-Class Laser Pulses onto Foam Targets},
  author = {Pugachev, L.P. and Andreev, N.E. and Levashov, P.R. and Rosmej, O.N.},
  year = {2016},
  month = sep,
  journal = {Nuclear Instruments and Methods in Physics Research Section A: Accelerators, Spectrometers, Detectors and Associated Equipment},
  volume = {829},
  pages = {88--93},
  issn = {01689002},
  doi = {10.1016/j.nima.2016.02.053},
  urldate = {2025-07-06},
  langid = {english}
}

@article{pukhovParticleAccelerationRelativistic1999,
  title = {Particle Acceleration in Relativistic Laser Channels},
  author = {Pukhov, A. and Sheng, Z.-M. and {Meyer-ter-Vehn}, J.},
  year = {1999},
  month = jul,
  journal = {Physics of Plasmas},
  volume = {6},
  number = {7},
  pages = {2847--2854},
  issn = {1070-664X, 1089-7674},
  doi = {10.1063/1.873242},
  urldate = {2025-03-17},
  langid = {english}
}

@article{rosmejAdvancedPlasmaTarget2025,
  title = {Advanced Plasma Target from Pre-Ionized Low-Density Foam for Effective and Robust Direct Laser Acceleration of Electrons},
  author = {Rosmej, Olga N. and Gyrdymov, Mikhail and Andreev, Nikolay E. and Tavana, Parysatis and Popov, Vyacheslav and Borisenko, Nataliya G. and Gromov, Alexandr I. and Gus'kov, Sergey Yu. and Yakhin, Rafael and Vegunova, Galina A. and Bukharskii, Nikolai and Korneev, Philipp and Cikhardt, Jakub and Z{\"a}hter, Sero and Busch, Sebastian and Jacoby, Joachim and Pimenov, Vladimir G. and Spielmann, Christian and Pukhov, Alexander},
  year = {2025},
  journal = {High Power Laser Science and Engineering},
  volume = {13},
  pages = {e3},
  issn = {2095-4719, 2052-3289},
  doi = {10.1017/hpl.2024.85},
  urldate = {2025-07-06},
  langid = {english}
}

@article{rosmejBrightBetatronRadiation2021a,
  title = {Bright Betatron Radiation from Direct-Laser-Accelerated Electrons at Moderate Relativistic Laser Intensity},
  author = {Rosmej, O. N. and Shen, X. F. and Pukhov, A. and Antonelli, L. and Barbato, F. and Gyrdymov, M. and G{\"u}nther, M. M. and Z{\"a}hter, S. and Popov, V. S. and Borisenko, N. G. and Andreev, N. E.},
  year = {2021},
  month = jul,
  journal = {Matter and Radiation at Extremes},
  volume = {6},
  number = {4},
  pages = {048401},
  issn = {2468-2047, 2468-080X},
  doi = {10.1063/5.0042315},
  urldate = {2025-07-06},
  langid = {english}
}

@article{rosmejHeatingLowdensityCHOfoam2011,
  title = {Heating of Low-Density {{CHO-foam}} Layers by Means of Soft {{X-rays}}},
  author = {Rosmej, O.N. and Bagnoud, V. and Eisenbarth, U. and Vatulin, V. and Zhidkov, N. and Suslov, N. and Kunin, A. and Pinegin, A. and Sch{\"a}fer, D. and Nisius, {\relax Th}. and Wilhein, {\relax Th}. and Rienecker, T. and Wiechula, J. and Jacoby, J. and Zhao, Y. and Vergunova, G. and Borisenko, N. and Orlov, N.},
  year = {2011},
  month = oct,
  journal = {Nuclear Instruments and Methods in Physics Research Section A: Accelerators, Spectrometers, Detectors and Associated Equipment},
  volume = {653},
  number = {1},
  pages = {52--57},
  issn = {01689002},
  doi = {10.1016/j.nima.2011.01.167},
  urldate = {2025-03-19},
  copyright = {https://www.elsevier.com/tdm/userlicense/1.0/},
  langid = {english}
}

@article{rosmejHighcurrentLaserdrivenBeams2020a,
  title = {High-Current Laser-Driven Beams of Relativistic Electrons for High Energy Density Research},
  author = {Rosmej, O N and Gyrdymov, M and G{\"u}nther, M M and Andreev, N E and Tavana, P and Neumayer, P and Z{\"a}hter, S and Zahn, N and Popov, V S and Borisenko, N G and Kantsyrev, A and Skobliakov, A and Panyushkin, V and Bogdanov, A and Consoli, F and Shen, X F and Pukhov, A},
  year = {2020},
  month = nov,
  journal = {Plasma Physics and Controlled Fusion},
  volume = {62},
  number = {11},
  pages = {115024},
  issn = {0741-3335, 1361-6587},
  doi = {10.1088/1361-6587/abb24e},
  urldate = {2025-07-06}
}

@article{rosmejHydrodynamicRadiativeProperties2015,
  title = {The Hydrodynamic and Radiative Properties of Low-Density Foams Heated by x-Rays},
  author = {Rosmej, O N and Suslov, N and Martsovenko, D and Vergunova, G and Borisenko, N and Orlov, N and Rienecker, T and Klir, D and Rezack, K and Orekhov, A and Borisenko, L and Krousky, E and Pfeifer, M and Dudzak, R and Maeder, R and Schaechinger, M and Schoenlein, A and Zaehter, S and Jacoby, J and Limpouch, J and Ullschmied, J and Zhidkov, N},
  year = {2015},
  month = sep,
  journal = {Plasma Physics and Controlled Fusion},
  volume = {57},
  number = {9},
  pages = {094001},
  issn = {0741-3335, 1361-6587},
  doi = {10.1088/0741-3335/57/9/094001},
  urldate = {2025-03-25},
  langid = {english}
}

@article{rosmejInteractionRelativisticallyIntense2019a,
  title = {Interaction of Relativistically Intense Laser Pulses with Long-Scale near Critical Plasmas for Optimization of Laser Based Sources of {{MeV}} Electrons and Gamma-Rays},
  author = {Rosmej, O N and Andreev, N E and Zaehter, S and Zahn, N and Christ, P and Borm, B and Radon, T and Sokolov, A and Pugachev, L P and Khaghani, D and Horst, F and Borisenko, N G and Sklizkov, G and Pimenov, V G},
  year = {2019},
  month = apr,
  journal = {New Journal of Physics},
  volume = {21},
  number = {4},
  pages = {043044},
  issn = {1367-2630},
  doi = {10.1088/1367-2630/ab1047},
  urldate = {2025-07-06}
}

@article{tajimaLaserElectronAccelerator1979a,
  title = {Laser {{Electron Accelerator}}},
  author = {Tajima, T. and Dawson, J. M.},
  year = {1979},
  month = jul,
  journal = {Physical Review Letters},
  volume = {43},
  number = {4},
  pages = {267--270},
  issn = {0031-9007},
  doi = {10.1103/PhysRevLett.43.267},
  urldate = {2025-07-05},
  copyright = {http://link.aps.org/licenses/aps-default-license},
  langid = {english}
}

@article{takabeRecentProgressLaboratory2021,
  title = {Recent Progress of Laboratory Astrophysics with Intense Lasers},
  author = {Takabe, Hideaki and Kuramitsu, Yasuhiro},
  year = {2021},
  journal = {High Power Laser Science and Engineering},
  volume = {9},
  pages = {e49},
  issn = {2095-4719, 2052-3289},
  doi = {10.1017/hpl.2021.35},
  urldate = {2025-07-05},
  langid = {english}
}

@article{tangInfluenceLaserFocusing2024a,
  title = {The Influence of Laser Focusing Conditions on the Direct Laser Acceleration of Electrons},
  author = {Tang, H and Tangtartharakul, K and Babjak, R and Yeh, I-L and Albert, F and Chen, H and Campbell, P T and Ma, Y and Nilson, P M and Russell, B K and Shaw, J L and Thomas, A G R and Vranic, M and Arefiev, A V and Willingale, L},
  year = {2024},
  month = may,
  journal = {New Journal of Physics},
  volume = {26},
  number = {5},
  pages = {053010},
  issn = {1367-2630},
  doi = {10.1088/1367-2630/ad3be4},
  urldate = {2025-07-06}
}

@article{tangtartharakulCollimatedGrayEmission2025,
  title = {Collimated {$\gamma$}-Ray Emission Enabled by Efficient Direct Laser Acceleration},
  author = {Tangtartharakul, K and Fauvel, G and Meir, T and Condamine, F P and Weber, S and Pomerantz, I and Manuel, M and Arefiev, A},
  year = {2025},
  month = feb,
  journal = {New Journal of Physics},
  volume = {27},
  number = {2},
  pages = {023024},
  issn = {1367-2630},
  doi = {10.1088/1367-2630/adb3c1},
  urldate = {2025-03-17},
  langid = {english}
}

@article{tavanaUltrahighEfficiencyBremsstrahlung2023,
  title = {Ultra-High Efficiency Bremsstrahlung Production in the Interaction of Direct Laser-Accelerated Electrons with High-{{Z}} Material},
  author = {Tavana, P. and Bukharskii, N. and Gyrdymov, M. and Spillmann, U. and Z{\"a}hter, {\c S}. and Cikhardt, J. and Borisenko, N. G. and Korneev, {\relax Ph}. and Jacoby, J. and Spielmann, C. and Andreev, N. E. and G{\"u}nther, M. M. and Rosmej, O. N.},
  year = {2023},
  month = may,
  journal = {Frontiers in Physics},
  volume = {11},
  pages = {1178967},
  issn = {2296-424X},
  doi = {10.3389/fphy.2023.1178967},
  urldate = {2025-03-17},
  langid = {english}
}

@article{wangQuasimonoenergeticLaserplasmaAcceleration2013a,
  title = {Quasi-Monoenergetic Laser-Plasma Acceleration of Electrons to 2 {{GeV}}},
  author = {Wang, Xiaoming and Zgadzaj, Rafal and Fazel, Neil and Li, Zhengyan and Yi, S. A. and Zhang, Xi and Henderson, Watson and Chang, Y.-Y. and Korzekwa, R. and Tsai, H.-E. and Pai, C.-H. and Quevedo, H. and Dyer, G. and Gaul, E. and Martinez, M. and Bernstein, A. C. and Borger, T. and Spinks, M. and Donovan, M. and Khudik, V. and Shvets, G. and Ditmire, T. and Downer, M. C.},
  year = {2013},
  month = jun,
  journal = {Nature Communications},
  volume = {4},
  number = {1},
  pages = {1988},
  issn = {2041-1723},
  doi = {10.1038/ncomms2988},
  urldate = {2025-07-05},
  langid = {english}
}

@article{willingaleSurfaceWavesElectron2013a,
  title = {Surface Waves and Electron Acceleration from High-Power, Kilojoule-Class Laser Interactions with Underdense Plasma},
  author = {Willingale, L and Thomas, A G R and Nilson, P M and Chen, H and Cobble, J and Craxton, R S and Maksimchuk, A and Norreys, P A and Sangster, T C and Scott, R H H and Stoeckl, C and Zulick, C and Krushelnick, K},
  year = {2013},
  month = feb,
  journal = {New Journal of Physics},
  volume = {15},
  number = {2},
  pages = {025023},
  issn = {1367-2630},
  doi = {10.1088/1367-2630/15/2/025023},
  urldate = {2025-07-06},
  copyright = {http://iopscience.iop.org/info/page/text-and-data-mining}
}

@article{wuXingGuangIIILaser2020,
  title = {{{XingGuang III}} Laser Facility and Its Experimental Ability to Drive High-Energy Particle Beams},
  author = {Wu, Yuchi and Zhu, Bin and Dong, Kegong and Lu, Feng and He, Shukai and Zhang, Bo and Yan, Yonghong and Yu, Minghai and Tan, Fang and Wang, Shaoyi and Zhang, Tiankui and Liu, DongXiao and Yang, Yue and Qian, Feng and Han, Dan and Zhou, Kainan and Zhao, Zongqing and Su, Jingqin and Cao, Leifeng and Zhou, Weimin and Gu, Yuqiu},
  year = {2020},
  month = sep,
  journal = {Laser Physics},
  volume = {30},
  number = {9},
  pages = {096001},
  issn = {1054-660X, 1555-6611},
  doi = {10.1088/1555-6611/aba3ca},
  urldate = {2025-03-25},
  langid = {english}
}

@article{zhangStudiesHighEnergy2016,
  title = {Studies of High Energy Density Physics and Laboratory Astrophysics Driven by Intense Lasers},
  author = {Zhang, J and Li, Y T and Chen, L M and Dong, Q L and Zhong, J Y and Wang, W M and Sheng, Z M and Zhao, G},
  year = {2016},
  month = may,
  journal = {Journal of Physics: Conference Series},
  volume = {717},
  pages = {012004},
  issn = {1742-6588, 1742-6596},
  doi = {10.1088/1742-6596/717/1/012004},
  urldate = {2025-07-05},
  copyright = {http://iopscience.iop.org/info/page/text-and-data-mining}
}

@article{gonsalvesPetawattLaserGuiding2019,
  title = {Petawatt {{Laser Guiding}} and {{Electron Beam Acceleration}} to 8 {{GeV}} in a {{Laser-Heated Capillary Discharge Waveguide}}},
  author = {Gonsalves, A. J. and Nakamura, K. and Daniels, J. and Benedetti, C. and Pieronek, C. and De Raadt, T. C. H. and Steinke, S. and Bin, J. H. and Bulanov, S. S. and Van Tilborg, J. and Geddes, C. G. R. and Schroeder, C. B. and T{\'o}th, {\relax Cs}. and Esarey, E. and Swanson, K. and {Fan-Chiang}, L. and Bagdasarov, G. and Bobrova, N. and Gasilov, V. and Korn, G. and Sasorov, P. and Leemans, W. P.},
  year = 2019,
  month = feb,
  journal = {Physical Review Letters},
  volume = {122},
  number = {8},
  pages = {084801},
  issn = {0031-9007, 1079-7114},
  doi = {10.1103/PhysRevLett.122.084801},
  urldate = {2025-12-04},
  langid = {english}
}

@article{esareyPhysicsLaserdrivenPlasmabased2009,
  title = {Physics of Laser-Driven Plasma-Based Electron Accelerators},
  author = {Esarey, E. and Schroeder, C. B. and Leemans, W. P.},
  year = 2009,
  month = aug,
  journal = {Reviews of Modern Physics},
  volume = {81},
  number = {3},
  pages = {1229--1285},
  issn = {0034-6861, 1539-0756},
  doi = {10.1103/RevModPhys.81.1229},
  urldate = {2026-06-21},
  copyright = {http://link.aps.org/licenses/aps-default-license},
  langid = {english}
}

@article{siegristSelffocusingPlasmaDue1976,
  title = {Self-Focusing in a Plasma Due to Ponderomotive Forces and Relativistic Effects},
  author = {Siegrist, M.R.},
  year = 1976,
  month = mar,
  journal = {Optics Communications},
  volume = {16},
  number = {3},
  pages = {402--407},
  issn = {00304018},
  doi = {10.1016/0030-4018(76)90028-6},
  urldate = {2026-06-21},
  copyright = {https://www.elsevier.com/tdm/userlicense/1.0/},
  langid = {english}
}

@article{sprangleRelativisticSelfFocusingShortPulse1987,
  title = {Relativistic {{Self-Focusing}} of {{Short-Pulse Radiation Beams}} in {{Plasmas}}},
  author = {Sprangle, P. and Tang, Cha-Mei and Esarey, E.},
  year = 1987,
  journal = {IEEE Transactions on Plasma Science},
  volume = {15},
  number = {2},
  pages = {145--153},
  issn = {0093-3813},
  doi = {10.1109/TPS.1987.4316677},
  urldate = {2026-06-21},
  copyright = {https://ieeexplore.ieee.org/Xplorehelp/downloads/license-information/IEEE.html}
}

@article{sunSelffocusingShortIntense1987,
  title = {Self-Focusing of Short Intense Pulses in Plasmas},
  author = {Sun, Guo Zheng and Ott, Edward and Lee, Y. C. and Guzdar, Parvez},
  year = 1987,
  month = feb,
  journal = {The Physics of Fluids},
  volume = {30},
  number = {2},
  pages = {526--532},
  issn = {0031-9171},
  doi = {10.1063/1.866349},
  urldate = {2026-06-21},
  langid = {english}
}

@article{sadighi-bonabiImprovingRelativisticSelffocusing2009,
  title = {Improving the Relativistic Self-Focusing of Intense Laser Beam in Plasma Using Density Transition},
  author = {{Sadighi-Bonabi}, R. and Habibi, M. and Yazdani, E.},
  year = 2009,
  month = aug,
  journal = {Physics of Plasmas},
  volume = {16},
  number = {8},
  pages = {083105},
  issn = {1070-664X, 1089-7674},
  doi = {10.1063/1.3202694},
  urldate = {2026-06-23},
  langid = {english}
}

@article{zengSelftruncatedIonizationInjection2014,
  title = {Self-Truncated Ionization Injection and Consequent Monoenergetic Electron Bunches in Laser Wakefield Acceleration},
  author = {Zeng, Ming and Chen, Min and Sheng, Zheng-Ming and Mori, Warren B. and Zhang, Jie},
  year = 2014,
  month = mar,
  journal = {Physics of Plasmas},
  volume = {21},
  number = {3},
  pages = {030701},
  issn = {1070-664X, 1089-7674},
  doi = {10.1063/1.4868404},
  urldate = {2026-06-23},
  langid = {english}
}

@article{zouEnhancedTargetNormal2014,
  title = {Enhanced Target Normal Sheath Acceleration Based on the Laser Relativistic Self-Focusing},
  author = {Zou, D. B. and Zhuo, H. B. and Yang, X. H. and Shao, F. Q. and Ma, Y. Y. and Yu, T. P. and Wu, H. C. and Yin, Y. and Ge, Z. Y. and Li, X. H.},
  year = 2014,
  month = jun,
  journal = {Physics of Plasmas},
  volume = {21},
  number = {6},
  pages = {063103},
  issn = {1070-664X, 1089-7674},
  doi = {10.1063/1.4882245},
  urldate = {2026-06-23},
  langid = {english}
}

@article{huangLargeangleStimulatedRaman2025,
  title = {Large-Angle Stimulated {{Raman}} Scattering Induced by Transverse Density Modulation},
  author = {Huang, Z. M. and Wang, Qing and Cheng, R. J. and Li, X. X. and Lv, S. Y. and Liu, D. J. and Xu, Z. Y. and Zhang, S. T. and Chen, Z. J. and Wang, Qiang and Xiao, C. Z. and Liu, Z. J. and Cao, L. H. and Zheng, C. Y. and He, X. T.},
  year = 2025,
  month = sep,
  journal = {Matter and Radiation at Extremes},
  volume = {10},
  number = {5},
  pages = {057403},
  issn = {2468-2047, 2468-080X},
  doi = {10.1063/5.0278141},
  urldate = {2026-06-23},
  langid = {english}
}

@article{lezhninParallelLaserBeam2024,
  title = {On Parallel Laser Beam Merger in Plasmas},
  author = {Lezhnin, K. V. and Qu, Kenan and Fisch, N. J. and Bulanov, S. V.},
  year = 2024,
  month = mar,
  journal = {Physics of Plasmas},
  volume = {31},
  number = {3},
  pages = {032114},
  issn = {1070-664X, 1089-7674},
  doi = {10.1063/5.0191094},
  urldate = {2026-06-23},
  langid = {english}
}

@article{guskovEquationStatePartially2023a,
  title = {Equation of State of a Partially Homogenized Plasma of Low-Dense Porous Matter},
  author = {Gus'kov, S. {\relax Yu}. and Yakhin, R. A.},
  year = 2023,
  month = jun,
  journal = {Physics of Plasmas},
  volume = {30},
  number = {6},
  pages = {062709},
  issn = {1070-664X, 1089-7674},
  doi = {10.1063/5.0145385},
  urldate = {2026-07-03},
  langid = {english}
}

@article{caoExperimentalStudyMedical2023a,
  title = {Experimental Study of Medical Isotopes{\textsuperscript{62,64}} {{Cu}} And{\textsuperscript{68}} {{Ga}} Production Using Intense Picosecond Laser Pulse},
  author = {Cao, Zongwei and Qi, Wei and Lan, Haoyang and Cui, Bo and Zhang, Xiaohui and Deng, Zhigang and Zhang, Zhimeng and Wang, Guanlin and Zhang, Liangqi and Li, Xiankai and Yuan, Yun and Li, Xinxiang and Zhao, Zongqing and Luo, Wen and Zhou, Weimin},
  year = 2023,
  month = may,
  journal = {Plasma Physics and Controlled Fusion},
  volume = {65},
  number = {5},
  pages = {055007},
  issn = {0741-3335, 1361-6587},
  doi = {10.1088/1361-6587/acc090},
  urldate = {2026-07-07}
}

@article{gyrdymovHighbrightnessBetatronEmission2024c,
  title = {High-Brightness Betatron Emission from the Interaction of a Sub Picosecond Laser Pulse with Pre-Ionized Low-Density Polymer Foam for {{ICF}} Research},
  author = {Gyrdymov, Mikhail and Cikhardt, Jakub and Tavana, Parysatis and Borisenko, Nataliya G. and Gus{\textasciiacute}kov, Sergey Yu. and Yakhin, Rafael A. and Vegunova, Galina A. and Wei, Wenqing and Ren, Jieru and Zhao, Yongtao and Hoffmann, Dieter H. H. and Deng, Zhigang and Zhou, Weimin and Cheng, Rui and Yang, Jie and Novotny, Jan and Shen, Xiaofei and Pukhov, Alexander and Jacoby, Joachim and Spielmann, Christian and Popov, Viacheslav S. and Veysman, Mikhail E. and Andreev, Nikolay E. and Rosmej, Olga N.},
  year = 2024,
  month = jun,
  journal = {Scientific Reports},
  volume = {14},
  number = {1},
  pages = {14785},
  issn = {2045-2322},
  doi = {10.1038/s41598-024-65490-7},
  urldate = {2026-07-07},
  langid = {english}
}

@article{gyrdymovUltraintensePulsedSource2026,
  title = {Ultra-Intense Pulsed Source of Ionizing Radiation Based on Direct Laser Acceleration of Electrons for Studying the {{FLASH}} Effect},
  author = {Gyrdymov, Mikhail and Bukharskii, Nikolai and Fabian, Vratislav and H{\"a}fner, Michael and Karoon, Pharewa and Borisenko, Nataliya G. and Cikhardt, Jakub and Z{\"a}hter, Sero and Korneev, Philipp and Jacoby, Joachim and Andreev, Nikolay E. and Rosmej, Olga N.},
  year = 2026,
  month = feb,
  journal = {Scientific Reports},
  volume = {16},
  number = {1},
  pages = {7164},
  issn = {2045-2322},
  doi = {10.1038/s41598-026-40281-4},
  urldate = {2026-07-07},
  langid = {english}
}

@article{pangPangProductionMedicalRadioisotopes512026a,
  title = {Production of Medical Radioisotopes{\textsuperscript{51}} {{Cr}},{\textsuperscript{62,64}} {{Cu}}, and{\textsuperscript{ 99 {\emph{m}} }} {{Tc}} by Laser-Induced Photonuclear Reactions*},
  author = {Pang, Xuan and Wu, Di and Sun, Bao-Hua and Wang, Mei-Zhi and Lan, Hao-Yang and Xia, Yu-Hui and Wang, Zhe-Nan and Xu, Xin-Lu and Yan, Xue-Qing},
  year = 2026,
  month = apr,
  journal = {Chinese Physics C},
  volume = {50},
  number = {4},
  pages = {044003},
  issn = {1674-1137, 2058-6132},
  doi = {10.1088/1674-1137/ae32fa},
  urldate = {2026-07-07}
}

@article{tavanaUltrahighFluxDirect2026,
  title = {Ultrahigh Flux of Direct Laser-Accelerated Electrons, {{MeV}} Photons, and Neutrons from Overdense Polymer Foams},
  author = {Tavana, Parysatis and Gyrdymov, Mikhail and Cikhardt, Jakub and Novotny, Jan and Kalla, Ren{\'e} and Boller, Pascal and K{\"u}hl, Thomas and Glorius, Jan and Spillmann, Uwe and Tentori, Alessandro and Kozlova, Ekaterina and Bukharskii, Nikolai and Spielmann, Christian and Rosmej, Olga N.},
  year = 2026,
  month = mar,
  journal = {Physical Review Applied},
  volume = {25},
  number = {3},
  pages = {034003},
  issn = {2331-7019},
  doi = {10.1103/5mpy-2jw5},
  urldate = {2026-07-07},
  langid = {english}
}

@article{yangExpressDiagnosticIntense2026a,
  title = {Express Diagnostic of Intense Laser-Driven {{MeV}} Radiation Source Using Copper Isotopes},
  author = {Yang, Mingzhe and Wang, Ziyao and Ren, Jieru and Wei, Wenqing and Chen, Benzheng and Ma, Bubo and Zhang, Shizheng and Liu, Lirong and Li, Fangfang and Xiong, Jie and Yue, Hongwei and Lai, Zeyu and Li, Wenxuan and Hoffmann, Dieter H.H. and Rosmej, Olga N. and Tawana, Parysatis and Andreev, N.E. and Umarov, I.R. and Deng, Zhigang and Qi, Wei and Wang, Shaoyi and Fan, Quanping and Yuan, Zongqiang and Wang, Weiwu and Cui, Bo and Zhang, Xiaohui and Wu, Yuchi and Zhou, Weimin and Su, Jingqin and Cheng, Rui and Zhao, Yongtao},
  year = 2026,
  month = apr,
  journal = {Nuclear Instruments and Methods in Physics Research Section A: Accelerators, Spectrometers, Detectors and Associated Equipment},
  volume = {1084},
  pages = {171188},
  issn = {01689002},
  doi = {10.1016/j.nima.2025.171188},
  urldate = {2026-07-07},
  langid = {english}
}

@article{litvak1970finite,
  title={Finite-amplitude wave beams in a magnetoactive plasma},
  author={Litvak, AG},
  journal={Sov. Phys. JETP},
  volume={30},
  number={344},
  pages={166},
  year={1970}
}

@article{maxSelfModulationSelfFocusingElectromagnetic1974,
  title = {Self-{{Modulation}} and {{Self-Focusing}} of {{Electromagnetic Waves}} in {{Plasmas}}},
  author = {Max, Claire Ellen and Arons, Jonathan and Langdon, A. Bruce},
  year = 1974,
  month = jul,
  journal = {Physical Review Letters},
  volume = {33},
  number = {4},
  pages = {209--212},
  issn = {0031-9007},
  doi = {10.1103/PhysRevLett.33.209},
  urldate = {2026-07-14},
  copyright = {http://link.aps.org/licenses/aps-default-license},
  langid = {english}
}
\end{document}